\documentclass[11pt]{article}

\usepackage[margin=1in]{geometry}
\usepackage{graphicx}
\graphicspath{{./}}
\usepackage{amsmath, amssymb}
\usepackage{booktabs}
\usepackage{float}
\usepackage[section]{placeins}
\usepackage[numbers,sort&compress]{natbib}
\usepackage[expansion=false]{microtype}
\usepackage{xurl}
\usepackage{setspace}
\usepackage{enumitem}
\usepackage[table]{xcolor}
\usepackage{multirow}
\usepackage[hidelinks]{hyperref}

\begin{document}

\title{Capacity Markets for Large Loads under Supply-Chain Constraints}

\author{Tong Liu\thanks{School of Electrical and Computer Engineering, Cornell University, Ithaca, NY 14853, USA. Corresponding author. Email: tl927@cornell.edu} \and Jacob Mays\thanks{School of Civil and Environmental Engineering, Cornell University, Ithaca, NY 14853, USA. Email: jacobmays@cornell.edu}}

\date{}

\maketitle
\thispagestyle{empty}

\begin{onehalfspace}
\begin{abstract}\normalsize
Motivated by the rapid growth of data centers, we develop a model to evaluate bring-your-own-capacity (BYOC) mandates and flexibility accreditation in capacity markets for new large loads with shared supply-chain constraints. With efficient pricing, BYOC mainly reallocates procurement between grid-built and self-built capacity and therefore has little welfare effect, while flexibility delivers a modest gain by reducing the effective capacity requirement. Under administrative price caps, mandates can improve static welfare by forcing data centers to internalize the full cost of capacity. The welfare ranking of the two instruments depends on supply-chain stress. At low or moderate stress, only the flexibility instrument raises welfare. Under severe stress with capped prices, the welfare gain from the BYOC obligation can exceed the gross flexibility benefit. The two instruments differ in their effects on a neighboring market: a unilateral BYOC mandate can crowd out its capacity investment, while flexibility produces essentially no spillover at our calibrated benchmark. Finally, applying current capacity non-performance penalties to flexible loads may lead to financial incentives that are too weak to induce truthful flexibility reporting.

\vspace{0.3cm}
\noindent\textbf{Keywords:} Capacity markets, large loads, data centers, supply-chain constraints, bring-your-own-capacity, flexibility accreditation, cross-market spillover

\end{abstract}
\end{onehalfspace}

\newpage
\doublespacing

\section{Introduction}

Surging large-load demand, predominantly from data centers, is colliding with constrained physical supply in U.S.\ electricity markets. In PJM, large-load growth (97\% from data centers) added over 5,000~MW of demand to the 2027/28 capacity auction, which cleared 6,623~MW below its reliability requirement~\citep{PJM_BRA_2027, PJM_PressRelease_2027}; PJM's 2025 long-term load forecast projects approximately 30~GW of data-center-driven peak load growth through 2030~\citep{PJM_LoadForecast_2025}. NYISO anticipates 1,400~MW of large load from data centers and semiconductor manufacturing by 2027~\citep{NYISO_RNA_2024, NYISO_GoldBook_2025}; its interconnection queue had grown to 29 large-load projects totaling roughly 6,055~MW by mid-2025~\citep{NYISO_LargeLoadQueue_2025}. But new generation takes years to deliver. Large power transformers alone require three- to five-year lead times~\citep{DOE_LPT_2024}, and interconnection queues add further delay~\citep{Gorman_Joule_2024, FERC_Order2023}.

The mismatch between the pace of new supply and demand has led to price shocks and reliability concerns in many U.S. systems, giving rise to a political consensus that regulators, market operators, and data center owners should limit the impact of data center entry on other electricity consumers. This consensus is summarized in the Ratepayer Protection Pledge, a White House initiative that promises ``to protect American consumers from price hikes due to data center energy and infrastructure requirements''~\citep{WhiteHouse_RatepayerPledge_2026}. In the context of capacity markets, the consensus has led to three types of interventions. The first is price caps or other changes to the administratively defined demand for capacity. The second is bring your own capacity (BYOC) requirements: large loads contract for or self-build generation to offset their reliability obligation. The third is flexible large load frameworks that reduce the capacity obligation of loads that can credibly curtail during scarcity~\citep{Walsh_FlexibleLoad_2026, Colangelo_GridInteractive_2025, Norris_RethinkingLoadGrowth_2025, Camus_FlexibleDataCenters_2025, Byers_Billimoria_2025}.

Capacity markets are designed to coordinate generation investment through price signals~\citep{Cramton_CapacityMarket_2006, Hobbs_GenExpansion_2007, Joskow_CapacityPayments_2008, Zuo_CapacityMarket_2025}. In the standard analysis, an increase in the capacity price caused by new demand is considered a pecuniary effect: unhedged loads pay more and incumbent suppliers earn more, but the higher price is also a signal through which scarcity is communicated and new investment is rewarded. The price increase alone therefore does not establish a ``cost causation'' rationale for intervention. Such an argument requires costs that are not already reflected in market prices, such as lumpy network upgrades needed to facilitate interconnection. Absent those costs, shielding incumbent load from the price effect is better understood as a distributional policy. In this sense, interventions may be considered a manifestation of the ``credibility trap,'' in which the prices necessary to attract investment trigger interventions that undermine those same prices~\citep{PJM_PoweringReliability_2026}. When policymakers are unwilling to sustain a market-wide scarcity price, BYOC allows the market to preserve an administratively capped price for existing load while imposing a separate procurement obligation on new large loads.

This paper analyzes the joint interaction of administrative price caps, BYOC mandates, and flexibility accreditation in systems with shared supply chain constraints. We build a two-market equilibrium model in which two capacity markets share a constrained equipment supply chain. Data centers choose how much load to bring onto the grid and may be required to self-build generation under a BYOC mandate. Their net load obligation depends on a flexibility credit that captures how much of their peak demand the system counts toward the reliability requirement. Market~A, calibrated to PJM, may impose a BYOC mandate and grant flexibility accreditation; Market~B approximates NYISO, where analogous interventions may emerge under a FERC June 2026 show-cause order~\citep{FERC_LargeLoadShowCause_2026}. The model separates the physical scarcity rent created by the shared equipment constraint from the visible capacity price and asks how the BYOC and flexibility policies affect load entry, reliability value, financial incidence, and neighboring-market investment.

Four main findings emerge. First, BYOC mandates have small welfare effects under efficient pricing, since grid-built and self-built capacity are close substitutes. Under administrative price caps, mandates can become a second-best correction because the mandate induces data centers to internalize the true cost of capacity rather than the suppressed clearing price. Second, the correction is targeted. It does not correct the welfare gap created by non-data-center load growth, and it can crowd out neighboring-market entry when the neighbor is also short of capacity. Third, at our flexibility benchmark with an 80\%-firm design, the welfare ranking of the two instruments is state dependent. At low or moderate supply-chain stress, the mandate has no welfare effect, as grid entry stays active and flexibility is the only lever that helps. Under severe stress, the mandate produces an effect that is comparable to or larger than that of flexibility. Fourth, flexibility is invisible in clearing prices when administrative caps bind, meaning that non-performance penalty designs, if applied to flexible loads, may not align incentives for truthful reporting: the expected penalty can be below the savings from underreporting. These welfare comparisons are static; the model does not characterize how segmenting procurement through BYOC may affect future market-based investment if it deepens the long-run credibility problem.

The paper proceeds as follows. Section~\ref{sec:model} develops the model and derives the cost conservation identity. Section~\ref{sec:calibrated} establishes the calibration and the baseline distortion between efficient and administrative clearing. Section~\ref{sec:byoc} analyzes the BYOC mandate as a second-best correction. Section~\ref{sec:flex} analyzes flexibility accreditation and its interaction with the mandate.
Section~\ref{sec:discussion} concludes.

\section{Model}
\label{sec:model}

\subsection{Environment and policy instruments}
\begin{figure}[!htbp]
    \centering
    \includegraphics[width=0.95\columnwidth]{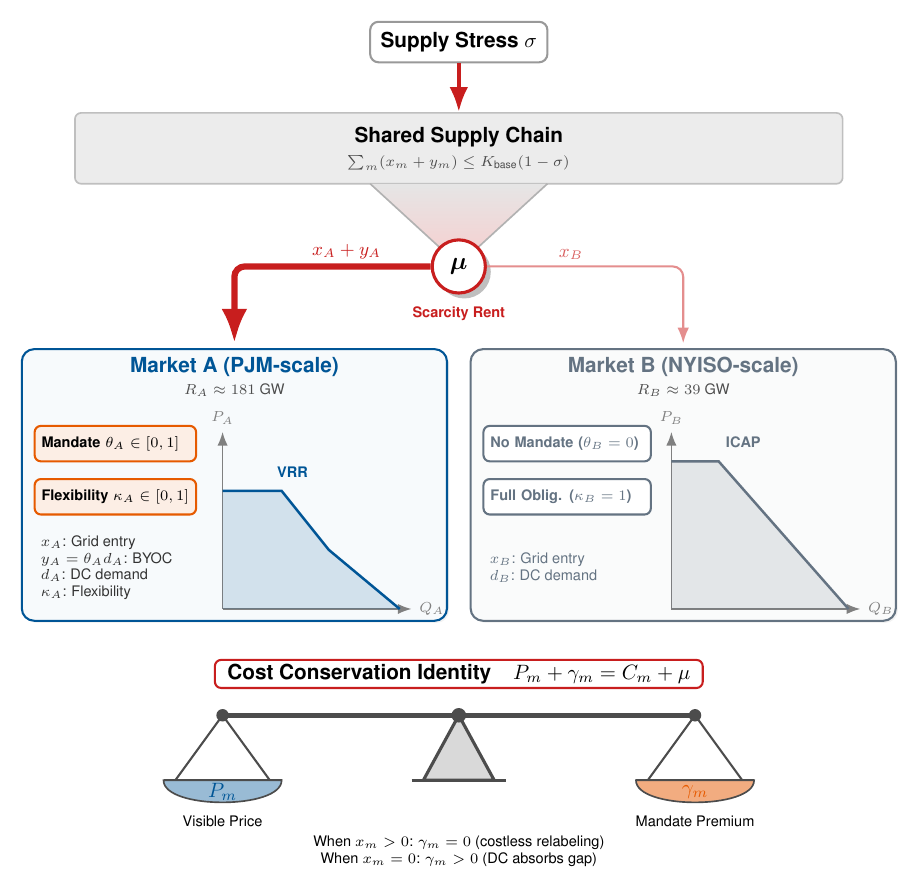}
    \caption{\textbf{Model structure.} Two capacity markets share an upstream supply chain (scarcity rent $\mu$). Market~A faces policy choices---mandate stringency $\theta_A$ and flexibility accreditation $\kappa_A$---while Market~B operates under business as usual ($\theta_B = 0$, $\kappa_B = 1$). The cost conservation identity $P_m + \gamma_m = C_m + \mu$ splits total procurement cost between the visible capacity price $P_m$ and the mandate premium $\gamma_m$.}
    \label{fig:mechanism}
\end{figure}

We model two neighboring capacity markets, $m \in \{A, B\}$, that run annual capacity auctions for a single generation technology and share a common upstream supply chain for physical generation equipment, such as turbines and transformers (Figure~\ref{fig:mechanism}). While in practice the markets would clear separately, the joint model represents a competitive equilibrium given the shared supply chain. Each market has an existing capacity stock $\text{ExCap}_m$, a reliability requirement $R_m$, and a levelized monthly cost of new entry $C_m$.

Three economic actors interact. Grid developers build $x_m \ge 0$~MW of standard new generation. Data centers bring $d_m \ge 0$~MW of new peak load, with $\bar{d}_m$ representing a forecasted large-load pipeline already embedded in the reliability requirement $R_m$. Under a BYOC mandate, data center developers also self-build $y_m$~MW of generation that counts toward the market's capacity balance. The mandate stringency is captured by $\theta_m$:
\begin{equation}
    y_m = \theta_m\, d_m, \qquad \theta_m \in [0,1],
    \label{eq:byoc}
\end{equation}
where $\theta_m = 0$ represents no mandate and $\theta_m = 1$ requires data centers to build one MW of generation for every MW of peak load they introduce. Both grid and self-built capacity draw from the same finite supply chain, where the parameter $\sigma$ represents increasing supply-chain stress:
\begin{equation}
    \sum_m (x_m + y_m) \le K_\text{base}(1 - \sigma), \qquad \sigma \in [0,1].
    \label{eq:supplychain}
\end{equation}
The shadow price of this constraint, $\mu \ge 0$, acts as a global scarcity rent for equipment access that transmits across both markets.

When a data center enters, it alters the local capacity balance. The system operator sets a static reliability requirement $R_m$ that already incorporates the forecasted large-load pipeline $\bar{d}_m$~\citep{PJM_BRA_2027, NYISO_RNA_2024, PJM_LoadForecast_2025}. Any forecasted demand that fails to materialize acts as a capacity buffer against this inflated target. The adjusted capacity position $Q_m$, evaluated by the demand curve via $z = Q_m / R_m$, decomposes as:
\begin{equation}
Q_m = \underbrace{\text{ExCap}_m}_{\text{existing capacity}} + \underbrace{x_m + y_m}_{\text{new build}} + \underbrace{(\bar{d}_m - d_m)}_{\text{unrealized load}} + \underbrace{(1 - \kappa_m)\, d_m}_{\text{flexibility credit}},\label{eq:capacity}
\end{equation}
where $\kappa_m \in [0,1]$ defines the data center's flexibility accreditation---the fraction of its peak load actually imposed on the grid as a capacity obligation~\citep{Bothwell_Hobbs_2017, Dvorkin_Silverman_2026}. The first three terms are the physical steel in the ground: existing capacity, grid entry ($x_m$), and self-builds ($y_m$, equal to $\theta_m d_m$ under the BYOC constraint~\eqref{eq:byoc}). The fourth term is the capacity buffer freed up when actual load obligations ($d_m$) fall short of the projected pipeline ($\bar{d}_m$). The final term is the flexibility credit, with realized data center load imposing an obligation of $\kappa_m\, d_m$ MW instead of its nameplate capacity.
Substituting $y_m = \theta_m d_m$ yields the equivalent reduced form $Q_m = \text{ExCap}_m + x_m + \bar{d}_m + (\theta_m - \kappa_m)\, d_m$.

Data centers derive private economic benefit from grid interconnection. We model the net value of entering market $m$ with demand $d_m$ as:
\begin{equation}
    V_m(d_m) = v_0\, d_m - \frac{v_{1,m}}{2}\, d_m^2, \qquad v_{1,m} = \frac{v_0}{\bar{d}_m},
    \label{eq:dcvalue}
\end{equation}
where the marginal value declines from $v_0$ to zero at the projection $\bar{d}_m$. The quadratic term captures diminishing returns to entry as successive data centers face less favorable sites and greater competition. The value of $v_0$ is calibrated to match observed capacity prices (Section~\ref{sec:calibration}).

\subsection{Clearing and welfare evaluation}

The market clearing problem maximizes global surplus under the assumed demand curve, including consumer reliability, data center value, and capacity costs:
\begin{equation}
    \max_{x, y, d} \;\; \sum_m \left[ \int_0^{Q_m} P_m(q)\, dq \;+\; V_m(d_m) \;-\; C_m(x_m + y_m) \right]
    \label{eq:objective}
\end{equation}
subject to (\ref{eq:byoc})--(\ref{eq:capacity}), $x_m \ge 0$, and $d_m \ge 0$. Capacity prices $P_m(q)$ are defined by administrative demand curves, such as PJM's Variable Resource Requirement (VRR) curve~\citep{PJM_Manual18_VRR, Brattle_VRR_2022} and NYISO's ICAP Demand Curve~\citep{NYISO_ICAP_Course_2025, NYISO2025Params}, which cap prices at an administrative maximum when reserve margins are low and decline to zero as they grow. Due to the administrative nature of these demand curves, the market clearing objective is not in general identical to a social welfare objective. As a social planner benchmark, we also consider Value of Lost Load (VOLL)-based clearing, where the demand curve is estimated based on an exponential function that tracks the probability of lost load as a function of the reserve margin:
\begin{equation} \label{eq:voll_exp}
D_m(z) = C_m \exp(-\beta_m(z - 1)),
\end{equation}
with $z = Q_m / R_m$~\citep{Zhao_Zheng_Litvinov_2018}.

We therefore separate the clearing mechanism from the welfare standard: the equilibrium is determined by each market's administrative demand curve, but welfare $W$ is evaluated using the VOLL integral, taken to be a better estimate of the true social cost of outages~\citep{Zhao_Zheng_Litvinov_2018}. We vary policy in Market~A only and fix Market~B to align with its current practice (i.e., $\theta_B = 0$ and $\kappa_B = 1$). We track two metrics throughout the analysis. The mandate cost measures the welfare change of requiring data centers to self-build their full capacity:
\begin{align}
    \text{mc} &= W(\theta_A{=}0,\, \kappa_A{=}1) \;-\; W(\theta_A{=}1,\, \kappa_A{=}1). \label{eq:mc}
\end{align}
The flexibility benefit measures the welfare gained when data centers are accredited as partly flexible, so the system counts 80\% of their peak demand as a capacity obligation and treats the remaining 20\% as curtailable:
\begin{align}
    \text{fb} &= W(\theta_A{=}0,\, \kappa_A{=}0.8) \;-\; W(\theta_A{=}0,\, \kappa_A{=}1). \label{eq:fb}
\end{align}

The 80\% level ($\kappa_A = 0.8$) is our benchmark: a data center that curtails 20\% of its peak draw during scarcity, matching the 80\% firm / 20\% conditional-firm design in recent industry proposals~\citep{Camus_FlexibleDataCenters_2025}. Appendix~\ref{app:robustness} traces the full $\kappa_A$ dependence, with $\kappa_A = 0.2$ (deep flexibility) as the aggressive end of the range. In the objective of Eq.~\eqref{eq:objective}, provision of flexibility is assumed to be costless; we discuss potential costs in Section~\ref{sec:provision}.

We also evaluate the firm-matched package $(\theta_A,\kappa_A)=(0.8,0.8)$, in which BYOC covers the portion of load that remains firm. The conditional comparisons (reported in Table~\ref{tab:part3}) are
\begin{align}
\text{fb}|_{\theta_A=0.8} &= W(\theta_A{=}0.8,\, \kappa_A{=}0.8) - W(\theta_A{=}0.8,\, \kappa_A{=}1), \label{eq:fb_cond}\\
\text{mc}|_{\kappa_A=0.8} &= W(\theta_A{=}0,\, \kappa_A{=}0.8) - W(\theta_A{=}0.8,\, \kappa_A{=}0.8), \label{eq:mc_cond}\\
\text{combined} &= W(\theta_A{=}0.8,\, \kappa_A{=}0.8) - W(\theta_A{=}0,\, \kappa_A{=}1). \label{eq:combined}
\end{align}

The combined effect decomposes as $\text{combined} = \text{fb} - \text{mc}|_{\kappa_A=0.8}$: the standalone flexibility benefit plus the mandate's marginal welfare contribution ($-\text{mc}|_{\kappa_A=0.8}$, positive where the mandate is welfare-improving) once flexibility is active. The full mandate cost mc ($\theta_A=1$) does not enter this decomposition.

Positive mc means the mandate reduces welfare; positive fb means flexibility improves it (before accounting for any provision costs). Throughout the results, $\theta$ and $\kappa$ refer to $\theta_A$ and $\kappa_A$ unless otherwise noted, as $\theta_B = 0$ and $\kappa_B = 1$ are fixed.

For each policy comparison, the welfare change decomposes into a consumer reliability term, a data center value term, and an investment cost term. Write $\Delta^{\text{mc}} f \equiv f(\theta_A{=}1, \kappa_A{=}1) - f(\theta_A{=}0, \kappa_A{=}1)$ and $\Delta^{\text{fb}} f \equiv f(\theta_A{=}0, \kappa_A{=}0.8) - f(\theta_A{=}0, \kappa_A{=}1)$ for any quantity $f$. Then

\begin{align}
\text{mc} &= -\sum_m \bigl(\Delta^{\text{mc}}\mathrm{Rel}_m + \Delta^{\text{mc}}\mathrm{DC}_m - \Delta^{\text{mc}}\mathrm{Inv}_m\bigr), \label{eq:mc_decomp} \\
\text{fb} &= \phantom{-}\sum_m \bigl(\Delta^{\text{fb}}\mathrm{Rel}_m + \Delta^{\text{fb}}\mathrm{DC}_m - \Delta^{\text{fb}}\mathrm{Inv}_m\bigr), \label{eq:fb_decomp}
\end{align}

where the three components are
\begin{equation}
\mathrm{Rel}_m = \int_{R_m}^{Q_m} D_m(q/R_m)\,dq, \qquad \mathrm{DC}_m = V_m(d_m), \qquad \mathrm{Inv}_m = C_m(x_m + y_m). \label{eq:decomp_terms}
\end{equation}
$\mathrm{Rel}_m$ is the gross load-side reliability value (or avoided outage cost) under the VOLL demand curve~\eqref{eq:voll_exp}, which values reliability at its true social value rather than the capped administrative price. $\mathrm{DC}_m$ is the gross data center private value from Eq.~\eqref{eq:dcvalue}. $\mathrm{Inv}_m$ is the capacity investment cost. This is the social planner welfare identity, while transfers between consumers, incumbent generators, new entrants, and data centers cancel in this sum and are reported separately in the distributional analysis (Section~\ref{sec:incidence}, Table~\ref{tab:distributional}).

\subsection{Cost conservation and data center entry}
\label{sec:kkt}
The equilibrium that determines these welfare metrics satisfies the following optimality conditions derived from~\eqref{eq:objective}.

Let $\mu \ge 0$ denote the dual of the supply-chain constraint, let $P_m\equiv P_m(Q_m)$ denote the marginal value on market $m$'s capacity demand curve, and let $\gamma_m$ be the dual of the BYOC equality constraint $y_m = \theta_m d_m$. The Lagrangian is:

\begin{align}
\mathcal{L} &= \sum_m \Bigl[ \int_0^{Q_m} P_m(q)\,dq + V_m(d_m) - C_m(x_m + y_m) \Bigr] \notag \\
&\quad - \mu \Bigl[\sum_m (x_m + y_m) - K_{\mathrm{eff}}\Bigr] + \sum_m \gamma_m (y_m - \theta_m d_m),
\label{eq:lagrangian}
\end{align}
where $K_{\mathrm{eff}} = K_{\mathrm{base}}(1-\sigma)$ and $Q_m = \mathrm{ExCap}_m + x_m + y_m + (\bar{d}_m - \kappa_m d_m)$ as in~\eqref{eq:capacity}, with $y_m$ entering both the integrand $P_m(Q_m)$ and the supply-chain constraint.

The stationarity condition on $x_m$ yields:
\begin{equation}
0 \le x_m \perp \bigl[C_m + \mu - P_m\bigr] \ge 0.
\label{eq:grid_foc}
\end{equation}
When grid entry occurs ($x_m > 0$), complementary slackness forces $P_m = C_m + \mu$. The clearing price exactly covers the construction cost plus the scarcity rent.

The stationarity condition on $y_m$ yields the cost conservation identity~\eqref{eq:byoc_foc}:
\begin{equation}
C_m + \mu = P_m + \gamma_m.
\label{eq:byoc_foc}
\end{equation}
Because $y_m = \theta_m d_m$ is an equality constraint, $\gamma_m$ is unrestricted in sign. When grid entry is positive ($x_m > 0$), substituting $P_m = C_m + \mu$ from \eqref{eq:grid_foc} immediately sets $\gamma_m = 0$. The mandate functions as a pure relabeling with no cost difference between procurement channels.

When grid entry is crowded out ($x_m = 0$), $P_m < C_m + \mu$, resulting in $\gamma_m = C_m + \mu - P_m > 0$. Under a mandate, this positive term is the premium data centers absorb to self-build capacity.

With $d_m > 0$, the stationarity condition on $d_m$ gives:
\begin{equation}
V'_m(d_m) = \theta_m \gamma_m + \kappa_m P_m.
\label{eq:dc_foc}
\end{equation}
where $V'_m(d_m) = v_0 - v_{1,m} d_m$ from \eqref{eq:dcvalue}. The left side is the marginal private value of entry. The right side aggregates two costs: the mandate cost ($\theta_m \gamma_m$, the premium on self-built capacity) and the net load obligation ($\kappa_m P_m$, the capacity price levied on the fraction of demand counting toward reliability).

Substituting the cost conservation identity \eqref{eq:byoc_foc} into \eqref{eq:dc_foc} yields:
\begin{equation}
V'_m(d_m) = \theta_m(C_m + \mu) + (\kappa_m - \theta_m) P_m.
\label{eq:dc_foc_expanded}
\end{equation}
Under a full gross-load obligation $(\theta_m,\kappa_m)=(1,1)$, the clearing price drops out and $V'_m=C_m+\mu$: entry responds to the constrained marginal cost rather than the administered market price. More generally, the firm-matched package $\theta_m=\kappa_m$ also eliminates the direct dependence on $P_m$, giving $V'_m=\theta_m(C_m+\mu)$.

Without a mandate ($\theta_m = 0$), the condition becomes $V'_m = \kappa_m P_m$. Entry depends on the clearing price and flexibility accreditation, while the mandate premium $\gamma_m$ drops out.

\section{Calibration and baseline results}
\label{sec:calibrated}

This section defines the calibrated markets, with Market~A based on PJM and Market~B representing a stressed NYISO-scale neighbor. Section~\ref{sec:calibration} fixes the calibration parameters using current PJM and NYISO regulatory filings. Section~\ref{sec:VOLL} derives the VOLL demand curve that defines the welfare benchmark from each market's reliability standards. Section~\ref{sec:baseline_vrrvoll} then quantifies the baseline distortion between efficient (VOLL) and administrative (VRR) clearing in the absence of any policy intervention.\footnote{For simplicity, we use the PJM-specific term VRR to apply to both markets.} That distortion is the welfare reference against which the BYOC mandate (Section~\ref{sec:byoc}) and flexibility accreditation (Section~\ref{sec:flex}) are measured.

\subsection{Calibration}
\label{sec:calibration}

\begin{table}[htbp]
\centering\small
\caption{Calibration parameters.}
\label{tab:params}
\begin{tabular*}{\textwidth}{@{\extracolsep{\fill}} l c c >{\raggedright\arraybackslash}p{0.42\textwidth} @{}}
\toprule
Parameter & Market A (PJM) & Market B & Source \\
\midrule
$R_m$ (MW) & 181,274 & 39,372 & Peak $\times$ (1+IRM)~\citep{PJM_RR, NYSRC_Peak_IRM} \\
$C_m$ (\$/MW-mo) & 5,500 & 5,720 & Net CONE~\citep{PJM_BRA_2025, NYISO2025Params} \\
ExCap$_m$ (MW) & 169,129 & 36,500 & See text;~\citep{PJM_BRA_2025,NYISO_GoldBook_2025} \\
$\bar{d}_m$ (MW)\footnotemark & 5,000 & 1,400 & \citep{PJM_BRA_2027, NYISO_RNA_2024, NYISO_GoldBook_2025} \\
$\beta_m$ & 35.0 & 17.7 & \citep{Brattle_VRR_2025, Potomac_SOM_2024} \\
LOLH$_{0,m}$ (h/yr) & 0.323 & 0.374 & \citep{PJM_PC_3IA_2025, NYSRC_Peak_IRM} \\
VOLL$_m$ (\$/MWh) & 213,070 & 191,375 & $12C_m / [\text{LOLH}_{0,m}(1{-}\text{EFORd}_m)]$~\citep{Zhao_Zheng_Litvinov_2018} \\
VRR/ICAP curve & \multicolumn{2}{c}{See note$^\dagger$} & \citep{PJM_Manual18_VRR, NYISO2025Params} \\
$v_0$ (\$/MW-mo) & \multicolumn{2}{c}{29{,}260} & Chosen so $P_A$ hits BRA cap ($\sigma=0.3$) \\
$v_{1,m}$ (\$/MW$^2$-mo) & 5.9 & 20.9 & $v_0/\bar{d}_m$ \\
$K_\text{base}$ (MW) & \multicolumn{2}{c}{8{,}311} & BRA shortfall (ICAP; see text)~\citep{PJM_BRA_2027} \\
Price cap (\$/MW-mo) & 8,250 & 21,690 & \citep{PJM_Manual18_VRR, NYISO2025Params} \\
\bottomrule
\end{tabular*}

\vspace{1ex}
\raggedright
\footnotesize $^\dagger$PJM VRR: three-point at $Q/R = (0.989, 1.016, 1.068)$, $P = (\$8{,}250, \$4{,}125, \$0)$/MW-mo; cap at $1.5 \cdot C_m$ binds when $Q_m/R_m \le 0.989$. NYISO ICAP: linear from $(\$21{,}690, Q/R{=}0.665)$ to $(\$0, Q/R{=}1.12)$.
\end{table}
\footnotetext{$\bar{d}_A$ is the forecast increment between PJM's 2026/27 and 2027/28 BRAs; $\bar{d}_B$ is NYISO's cumulative large-load forecast for 2027. Sensitivity to $\bar{d}_B$ is examined in Appendix~\ref{app:shapley}.}

Table~\ref{tab:params} summarizes the calibration parameters. We anchor $\text{ExCap}_A = 0.933 \times R_A$ to PJM's 2025/26 BRA cleared-capacity ratio~\citep{PJM_BRA_2025}. With $\text{ExCap}_A$ fixed at this value, $v_0$ is calibrated to the price target, chosen such that $P_A$ reaches the PJM VRR cap ($8{,}250 = 1.5\cdot C_A$~\citep{PJM_Manual18_VRR}) at $\sigma = 0.3$. For Market~B, we set ExCap$_B = 36{,}500$~MW, below the 2025 NYISO Gold Book baseline ($\sim$40,910~MW)~\citep{NYISO_GoldBook_2025} but consistent with near-term reliability concerns connected to anticipated generation retirements~\citep{NYISO_CRP_2024}. We formulate and solve all models in Julia using JuMP.jl~\citep{Lubin_JuMP_2023} and Gurobi 13, linearizing the nonlinear demand curves with 1,000 segments.

Two calibration choices warrant note. First, $\bar{d}_A = 5{,}000$~MW (from PJM's 2027/28 BRA forecast increment) is applied against $R_A = 181{,}274$~MW (the 2025/26 reliability requirement); this reflects forward-looking demand against present-day supply, treating the calibrated case as a near-term scenario rather than a steady-state equilibrium. Second, ExCap$_B = 36{,}500$~MW is chosen to make cross-market spillovers visible: with surplus capacity in Market~B, there would be no spillover from A's mandate. Because the relevant equipment supply chains are national or global, the spillover results are best interpreted as an illustration of cross-jurisdictional incidence, not as a point estimate for NYISO.

We use $K_\text{base} = 8{,}311$~MW for the scarcity calibration, equal to the PJM 2027/28 BRA capacity shortfall converted to ICAP (6{,}623~MW UCAP divided by the pool-wide Accredited UCAP factor of 0.7969~\citep{PJM_RR}). Accordingly, $\sigma = 0$ is the least-tight point in an already constrained benchmark rather than a no-scarcity case; higher values of $\sigma$ can be interpreted as further tightening around this near-term shortage. Section~\ref{sec:slack} reports a counterfactual in which the constraint is relaxed. The chosen value of $K_\text{base}$ is not a direct measurement of upstream equipment supply constraints (e.g., turbine OEM throughput or transformer lead times); mapping to such data would require independent calibration outside our scope.

\paragraph{Comparison to forward-looking Net CONE estimates.}
Our $C_A = \$5{,}500$/MW-month uses the PJM 2025/26 BRA Net
CONE, matching the delivery year of $R_A$ and the VRR curve.
We retain this value as the administrative Net CONE anchor because the model separately represents short-run supply-chain scarcity through the endogenous rent \(\mu\): the effective marginal cost of capacity is \(C_A+\mu\). Recent forward-looking studies report substantially higher entry costs for later delivery years, including 2028/29 Gross CONE values of \(\$663/\text{MW-day}\) (\$20{,}169/MW-month) for a combustion turbine and \(\$813/\text{MW-day}\) (\$24{,}729/MW-month) for a combined cycle, as well as PJM combined-cycle project filings of roughly \(\$2.1\)--\(\$2.3~\text{M}/\text{MW}\) \citep[pp.~19--22, citing Brattle 2025 and GridLab 2025] {PJM_PoweringReliability_2026}. These estimates suggest that constrained entry costs can exceed legacy Net CONE anchors, but they are gross cost estimates and may already reflect equipment, EPC, and financing scarcity. We therefore interpret them as evidence on \(C_A+\mu\) or on supply-chain tightness rather than using them as direct replacements for~\(C_A\).

\subsection{VOLL demand curve} \label{sec:VOLL}
The VOLL demand curve in Eq.~\eqref{eq:voll_exp} is calibrated in
two steps. While this subsection offers a summary, full derivations, sensitivity analysis, and
discussion of forward versus reverse calibration methods appear in Appendix~\ref{app:voll_derivation}. First, the level is set by equating the marginal reliability value of capacity at the target reserve margin to annualized Net CONE, following the marginal reliability
framework in~\citet{Zhao_Zheng_Litvinov_2018}. The marginal reliability identity is
$dEUE/dQ = -\mathrm{LOLH} \cdot (1-\mathrm{EFORd})$, yielding
$\mathrm{VOLL}_A = \$213{,}070$/MWh and $\mathrm{VOLL}_B = \$191{,}375$/MWh,
both close to the \$216{,}000 ISO-NE estimate in the same reference.

Second, the slope $\beta$, which controls how steeply reliability degrades below the target reserve margin, is calibrated from LOLE data using the
exponential form $\mathrm{LOLE}(z) = \mathrm{LOLE}_0 \exp(-\beta(z-1))$.
For PJM, we fit the LOLE curve in PJM PRISM~\citep[Fig.~27]{Brattle_VRR_2025}
across $z \in [0.96, 1.06]$, yielding $\beta_A = 35$. For NYISO, the
forward MRI method $\beta_B = \mathrm{MRI} \cdot R / \mathrm{LOLE}_0$
applied to the reported MRI of 0.0052~days/100~MW~\citep{Potomac_SOM_2024}
yields $\beta_B = 17.7$.

Both markets are calibrated on LOLE data for consistency with the
1-in-10 reliability standard. The demand-curve formula formally uses
LOLH, but $\mathrm{LOLE}_0$ and $\mathrm{LOLH}_0$ share the same
exponential structure up to an event-duration factor; using
$\beta_{\mathrm{LOLE}}$ in place of $\beta_{\mathrm{LOLH}}$ yields a
shallower demand curve and conservative welfare results.

\subsection{Baseline VOLL vs VRR clearing}
\label{sec:baseline_vrrvoll}

Before introducing any policy lever, we characterize the baseline distortion between efficient (VOLL) and administrative (VRR) clearing at $\theta_A = 0$ and $\kappa_A = 1$. Table~\ref{tab:baseline_vrrvoll} reports prices in both markets and the implied aggregate welfare gap, $W^\text{VOLL} - W^\text{VRR}$.

\begin{table}[H]
\centering\small
\caption{Baseline VOLL vs VRR clearing under no policy ($\theta_A = 0$, $\kappa_A = 1$). Prices in \$/MW-month; welfare gap in \$M/month. Both markets' VRR prices stay flat while VOLL prices climb with $\sigma$; the rightmost column is the resulting aggregate welfare distortion.}
\label{tab:baseline_vrrvoll}
\begin{tabular}{@{}rrrrrr@{}}
\toprule
$\sigma$ & $P_A^\text{VRR}$ & $P_A^\text{VOLL}$ & $P_B^\text{VRR}$ & $P_B^\text{VOLL}$ & gap \\
\midrule
0.0 & $8{,}250$ & $10{,}148$ & $8{,}461$ & $10{,}362$ & $+3.06$ \\
0.3 & $8{,}250$ & $13{,}104$ & $8{,}461$ & $13{,}260$ & $+3.55$ \\
0.5 & $8{,}250$ & $15{,}488$ & $8{,}461$ & $15{,}072$ & $+9.00$ \\
0.7 & $8{,}250$ & $18{,}859$ & $8{,}461$ & $15{,}072$ & $+20.39$ \\
0.9 & $8{,}250$ & $23{,}040$ & $8{,}461$ & $15{,}072$ & $+40.68$ \\
\bottomrule
\end{tabular}
\end{table}

\begin{table}[H]
\centering\small
\caption{Baseline entry quantities under VRR vs VOLL clearing ($\theta_A = 0$, $\kappa_A = 1$). $d_m$ = data center entry, $x_m$ = grid entry, all in MW. Under VRR clearing, DC entry is pinned across $\sigma$; under VOLL, DC entry responds to stress and Market~B's grid entry collapses at $\sigma \ge 0.5$, mirroring the price contrast in Table~\ref{tab:baseline_vrrvoll}.}
\label{tab:baseline_quantities}
\begin{tabular}{@{}rrrrrrrrr@{}}
\toprule
& \multicolumn{2}{c}{$d_A$ (MW)} & \multicolumn{2}{c}{$d_B$ (MW)} & \multicolumn{2}{c}{$x_A$ (MW)} & \multicolumn{2}{c}{$x_B$ (MW)} \\
\cmidrule(lr){2-3}\cmidrule(lr){4-5}\cmidrule(lr){6-7}\cmidrule(lr){8-9}
$\sigma$ & VRR & VOLL & VRR & VOLL & VRR & VOLL & VRR & VOLL \\
\midrule
0.0 & 3{,}590 & 3{,}280 & 995 & 908 & 8{,}108 & 7{,}253 & 203 & 1{,}058 \\
0.3 & 3{,}590 & 2{,}795 & 995 & 772 & 5{,}615 & 5{,}444 & 203 &     374 \\
0.5 & 3{,}590 & 2{,}372 & 995 & 683 & 3{,}953 & 4{,}156 & 203 &       0 \\
0.7 & 3{,}590 & 1{,}730 & 995 & 683 & 2{,}291 & 2{,}493 & 203 &       0 \\
0.9 & 3{,}590 & 1{,}105 & 995 & 683 &     628 &     831 & 203 &       0 \\
\bottomrule
\end{tabular}
\end{table}

Table~\ref{tab:baseline_quantities} compares baseline entry under both clearing rules. Under VRR clearing, the cap locks $P_A$ at \$8{,}250 and $P_B$ at \$8{,}461 across all $\sigma$. The DC entry first-order condition $d_m = (v_0 - \kappa_m P_m)/v_{1,m}$ then pins $d_A$ at 3{,}590~MW and $d_B$ at 995~MW. Grid entry $x_A$ absorbs the supply-chain tightening, falling from 8{,}108 to 628~MW.

Under VOLL clearing, the efficient price rises with $\sigma$, raising the marginal capacity cost data centers face. Data center entry now responds to stress: $d_A$ falls from 3{,}280 to 1{,}105~MW and $d_B$ from 908 to 683~MW. Market~B's grid entry also responds, with $x_B^{\text{VOLL}}$ shifting from 1{,}058~MW at low stress to zero at $\sigma \ge 0.5$ as the shared supply chain tightens. Figure~\ref{fig:entry_supply} in Section~\ref{sec:flex} extends this comparison to multiple policy scenarios.

The welfare gap grows from \$3 to \$41~M/month as supply-chain stress tightens, driven primarily by Market~A's binding cap (Market~B's VRR clearing is closer to its VOLL benchmark). This baseline establishes the distortion that the policy instruments in the following sections will be measured against.

\section{BYOC mandates}
\label{sec:byoc}

We now turn to the policy lever of BYOC mandates ($\theta_A = 1$), which force data centers to procure supply that counts toward the market's capacity balance. Throughout this section we hold $\kappa_A = 1$: data centers' full peak load counts toward the reliability requirement, and the only policy change is whether they are required to self-build generation. At our moderate-stress baseline ($\sigma = 0.3$), the no-policy equilibrium yields $d_A \approx 3{,}590$~MW of data center entry and $x_A = 5{,}615$~MW of grid generation in Market~A, with $d_B \approx 995$~MW and $x_B = 203$~MW in Market~B; these quantities reappear throughout the analysis. This isolation lets us trace the mandate's aggregate welfare effect (Section~\ref{sec:second_best}), how it distributes across consumers, data centers, and generators (Section~\ref{sec:incidence}), how it spills into the neighboring market through the shared supply chain (Section~\ref{sec:spillovers}), and its limits under non-data-center load growth (Section~\ref{sec:nondc_byoc}).

\subsection{Second-best correction under the VRR cap}
\label{sec:second_best}

\begin{figure}[H]
    \centering
    \includegraphics[width=0.5\textwidth]{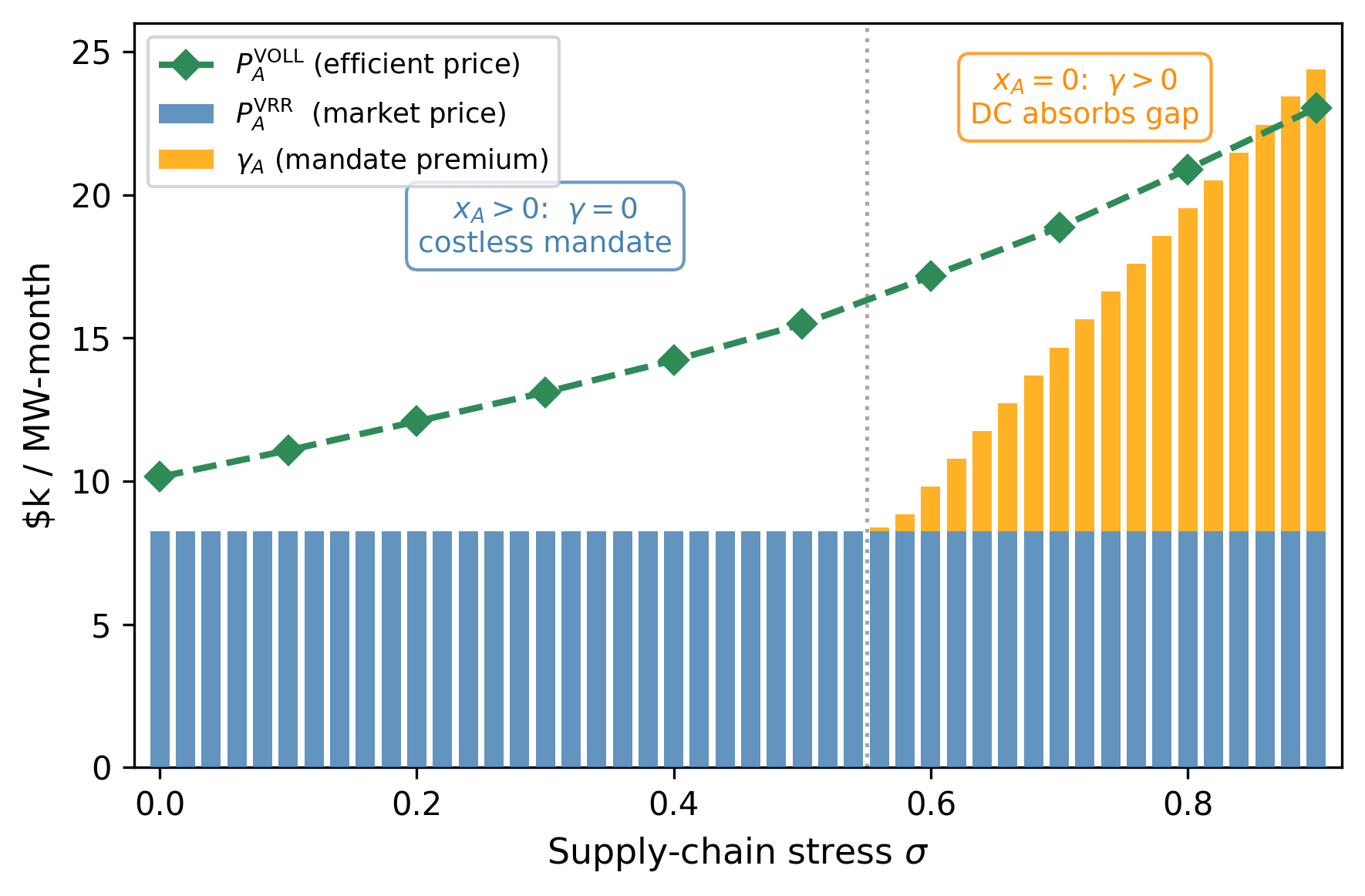}
    \caption{\textbf{Cost conservation under mandate} ($\theta_A = 1$, $\kappa_A = 1$).  The stacked bars show the identity $P_A^{\mathrm{VRR}}+\gamma_A=C_A+\mu$ within the VRR equilibrium. The green dashed line is the clearing price from a separate VOLL counterfactual and is shown as an efficient-pricing benchmark.}
    \label{fig:cost_cons}
\end{figure}

The mandate's welfare effect depends on the clearing rule. Under efficient VOLL pricing, the mandate is a pure relabeling whenever grid entry remains active. If $x_A > 0$, the cost conservation identity~\eqref{eq:byoc_foc} forces $\gamma_A = 0$, so requiring $y_A = d_A$ shifts capacity between grid and self-built channels at the same marginal cost, and welfare is unchanged (mc $= 0$); we call this \textbf{Regime~1}. Under VRR clearing, the cap binds in Market~A and $P_A$ stays at \$8{,}250 even when the efficient price would exceed it. The mandate then becomes a second-best correction: it forces data centers to face the true capacity cost $C_A + \mu$ through the mandate premium $\gamma_A = C_A + \mu - P_A$, bypassing the suppressed clearing price. Figure~\ref{fig:cost_cons} traces the decomposition. As $\sigma$ rises and grid entry is crowded out, $\gamma_A$ rises sharply: data centers absorb the gap between the cap and the true cost; we call this \textbf{Regime~2}.

Table~\ref{tab:mc_vrrvoll} quantifies the second-best correction. Under VOLL clearing, mc is zero across $\sigma \le 0.7$; the small residual at $\sigma = 0.9$ (+0.32~\$M/month) reflects supply-chain tightness even at efficient pricing. Under VRR, mc turns negative at $\sigma \ge 0.7$, reaching $-39$~\$M/month at $\sigma = 0.9$. Negative mc indicates the mandate raises welfare by disciplining load entry when the visible price is suppressed. The residual gap column shows the welfare loss the mandate alone cannot close: it shrinks from \$9~M/month at $\sigma = 0.5$ to \$1.30~M/month at $\sigma = 0.9$, but does not vanish.

\begin{table}[H]
\centering\small
\caption{Mandate cost (mc) under VRR vs VOLL clearing and residual welfare gap at $\theta_A{=}1$, $\kappa_A{=}1$. Negative mc means the mandate raises welfare (a second-best correction). Shaded rows ($\sigma \ge 0.7$) correspond to Regime~2 (grid entry crowded out). All welfare values in \$M/month.}
\label{tab:mc_vrrvoll}
\begin{tabular}{@{}rrrr@{}}
\toprule
$\sigma$ & mc$^\text{VRR}$ & mc$^\text{VOLL}$ & gap \\
\midrule
\rowcolor{gray!12} 0.3 & $0$ & $0$ & $3.55$ \\
\rowcolor{gray!12} 0.5 & $0$ & $0$ & $9.00$ \\
\rowcolor{gray!30} 0.7 & $\mathbf{-16.48}$ & $0$ & $3.91$ \\
\rowcolor{gray!30} 0.9 & $\mathbf{-39.06}$ & $+0.32$ & $1.30$ \\
\bottomrule
\end{tabular}
\end{table}

\subsection{Welfare decomposition and financial incidence}
\label{sec:incidence}

The mandate's effect is concentrated in Market~A (mc$_A$ from $0$ to $-41.17$~\$M); Market~B's contribution is smaller in our calibration (mc$_B \le +2.2$~\$M) and operates indirectly via the shared scarcity rent. Table~\ref{tab:distributional} reports the decomposition into consumer reliability value, data center value, and investment cost terms, market by market.

\begin{table}[htbp]
\centering\small
\caption{Welfare decomposition of the BYOC mandate by market. $\Delta\mathrm{Rel}$ is gross load-side reliability value, $\Delta\mathrm{DC}$ is gross data-center entry value, and $\Delta\mathrm{Inv}$ is investment cost. Shading denotes Market~A's regime. Market~B incurs a positive cost from the mandate as generation investment is instead directed to Market~A. Welfare values in \$M/month.}
\label{tab:distributional}
\begin{tabular}{@{}rrrrr@{}}
\toprule
$\sigma$ & $\Delta$Rel & $\Delta$DC & $\Delta$Inv & mc \\
\midrule
\multicolumn{5}{@{}l}{\textit{Market A}} \\
\rowcolor{gray!12} 0.0 & $0$ & $0$ & $0$ & $0$ \\
\rowcolor{gray!12} 0.3 & $0$ & $0$ & $0$ & $0$ \\
\rowcolor{gray!12} 0.5 & $0$ & $0$ & $0$ & $0$ \\
\rowcolor{gray!30} 0.7 & $+32.28$ & $-12.57$ & $+1.12$ & $-18.59$ \\
\rowcolor{gray!30} 0.9 & $+87.32$ & $-45.04$ & $+1.12$ & $-41.17$ \\
\addlinespace
\multicolumn{5}{@{}l}{\textit{Market B}} \\
\rowcolor{gray!12} 0.0 & $0$ & $0$ & $0$ & $0$ \\
\rowcolor{gray!12} 0.3 & $0$ & $0$ & $0$ & $0$ \\
\rowcolor{gray!12} 0.5 & $0$ & $0$ & $0$ & $0$ \\
\rowcolor{gray!30} 0.7 & $-3.19$ & $-0.08$ & $-1.16$ & $+2.12$ \\
\rowcolor{gray!30} 0.9 & $-3.19$ & $-0.08$ & $-1.16$ & $+2.12$ \\
\bottomrule
\end{tabular}
\end{table}

At moderate stress ($\sigma \le 0.5$), the mandate is a pure relabeling: all channels are zero in both markets. From $\sigma = 0.7$ onward, the mandate produces parallel trade-offs in both markets, linked through the shared scarcity rent $\mu$. In Market~A, the mandate forces $y_A = d_A$ onto a saturated supply chain, pinning $d_A$ at $K_{\mathrm{eff}}$ (below its $x_A > 0$ level of 3,590~MW). Forced BYOC raises consumer reliability ($\Delta$Rel$_A$ from $+32.28$ to $+87.32$~\$M/mo) at the cost of compressed DC value ($\Delta$DC$_A$ from $-12.57$ to $-45.04$~\$M/mo); mc$_A$ deepens from $-18.59$ to $-41.17$~\$M/mo.

The saturated chain in turn pushes $\mu$ up sharply, crowding out B's grid entry ($x_B \to 0$); B's consumer reliability drops ($\Delta$Rel$_B = -3.19$~\$M/mo) and B's investment falls ($\Delta$Inv$_B = -1.16$~\$M/mo, from foregone $x_B$), yielding mc$_B = +2.12$~\$M/mo. The nearly mirror-image $\Delta$Inv values ($+1.12$ in A, $-1.16$ in B) reflect supply-chain capacity shifting from B's $x_B$ to A's $y_A$.

Table~\ref{tab:transfer} highlights the distributional implications of the alternative designs. Because the capacity price applies to the full incumbent capacity base, restoring VOLL clearing would substantially increase payments by load and revenues to incumbent generators even when the immediate objective is to inform marginal entry. Conversely, the VRR cap produces a transfer on the existing-capacity base from incumbent generators to load. This transfer does not enter social welfare, but its magnitude helps explain the political pressure to suppress the common market price.
Table~\ref{tab:transfer} reports the implied monthly transfer,
$(P_m^{\mathrm{VOLL}} - P_m^{\mathrm{VRR}}) \cdot \mathrm{ExCap}_m$;
positive values mean generators receive less than under the VOLL counterfactual.

\begin{table}[H]
\centering\small
\caption{Incumbent revenue difference under VRR and VOLL clearing.
Transfer $= (P^{\mathrm{VOLL}} - P^{\mathrm{VRR}}) \cdot \mathrm{ExCap}_m$;
positive means incumbent generators receive less under VRR. Prices in
\$/MW-mo; transfers in \$M/mo. The transfer grows with $\sigma$.
The third column is the firm-matched policy ($\theta_A = 0.8$, $\kappa_A = 0.8$).}
\label{tab:transfer}
\begin{tabular}{@{}r|rrr|rrr|rrr@{}}
\toprule
& \multicolumn{3}{c|}{$\theta_A{=}0,\;\kappa_A{=}1$}
& \multicolumn{3}{c|}{$\theta_A{=}1,\;\kappa_A{=}1$}
& \multicolumn{3}{c}{$\theta_A{=}0.8,\;\kappa_A{=}0.8$} \\
\cmidrule(lr){2-4} \cmidrule(lr){5-7} \cmidrule(lr){8-10}
$\sigma$ & $P^{\mathrm{VRR}}$ & $P^{\mathrm{VOLL}}$ & Tr
& $P^{\mathrm{VRR}}$ & $P^{\mathrm{VOLL}}$ & Tr
& $P^{\mathrm{VRR}}$ & $P^{\mathrm{VOLL}}$ & Tr \\
\midrule
\multicolumn{10}{@{}l}{\textit{Market A}} \\
0.0 & $8{,}250$ & $10{,}148$ & $+321$ & $8{,}250$ & $10{,}148$ & $+321$ & $8{,}250$ & $9{,}713$ & $+247$ \\
0.3 & $8{,}250$ & $13{,}104$ & $+821$ & $8{,}250$ & $13{,}104$ & $+821$ & $8{,}250$ & $12{,}629$ & $+741$ \\
0.5 & $8{,}250$ & $15{,}488$ & $+1{,}224$ & $8{,}250$ & $15{,}488$ & $+1{,}224$ & $8{,}250$ & $15{,}326$ & $+1{,}197$ \\
0.7 & $8{,}250$ & $18{,}859$ & $+1{,}794$ & $8{,}250$ & $18{,}859$ & $+1{,}794$ & $8{,}250$ & $19{,}518$ & $+1{,}906$ \\
0.9 & $8{,}250$ & $23{,}040$ & $+2{,}501$ & $8{,}250$ & $21{,}853$ & $+2{,}301$ & $8{,}250$ & $21{,}853$ & $+2{,}301$ \\
\addlinespace
\multicolumn{10}{@{}l}{\textit{Market B}} \\
0.0 & $8{,}461$ & $10{,}362$ & $+69$ & $8{,}461$ & $10{,}362$ & $+69$ & $8{,}461$ & $9{,}945$ & $+54$ \\
0.3 & $8{,}461$ & $13{,}260$ & $+175$ & $8{,}461$ & $13{,}260$ & $+175$ & $8{,}461$ & $13{,}182$ & $+172$ \\
0.5 & $8{,}461$ & $15{,}072$ & $+241$ & $8{,}461$ & $15{,}072$ & $+241$ & $8{,}461$ & $15{,}072$ & $+241$ \\
0.7 & $8{,}461$ & $15{,}072$ & $+241$ & $8{,}695$ & $15{,}072$ & $+233$ & $8{,}695$ & $15{,}072$ & $+233$ \\
0.9 & $8{,}461$ & $15{,}072$ & $+241$ & $8{,}695$ & $15{,}072$ & $+233$ & $8{,}695$ & $15{,}072$ & $+233$ \\
\bottomrule
\end{tabular}
\end{table}

\subsection{Cross-market spillovers}
\label{sec:spillovers}

Section~\ref{sec:incidence} showed Market~B bearing a small welfare cost from A's mandate, transmitted through the shared scarcity rent. Here we trace the underlying mechanism: when does A's mandate begin to crowd out B's grid investment? At the baseline (ExCap$_B = 36{,}500$~MW, just below $R_B = 39{,}372$~MW), Market~B's deficit position supports 203~MW of new grid entry at $\sigma = 0$ ($x_B > 0$ with $P_B = \$8{,}461 \approx C_B + \mu$). This entry persists at low stress under a Market~A  mandate. As $\sigma$ rises, A's mandate consumes more of the shared supply chain and pushes $\mu$ up, eventually making B's grid entry unprofitable ($x_B \to 0$, Table~\ref{tab:spillover_detail}). The crowding-out emerges at $\sigma \ge 0.7$ in the baseline calibration; pushing ExCap$_B$ below 36,000~MW shifts this threshold to lower $\sigma$ values.

\begin{table}[H]
\centering\small
\caption{Equilibrium at the baseline ExCap$_B = 36{,}500$~MW. At $\sigma \le 0.5$, $x_B = 203$~MW under both base and mandate. At $\sigma \ge 0.7$, the mandate crowds out $x_B$ entirely. Figure~\ref{fig:phase}(a) maps the crowding out across the full ($\sigma$, ExCap$_B$) space. All quantities in MW.}
\label{tab:spillover_detail}
\begin{tabular}{@{}r|rrrr|rrrr@{}}
\toprule
& \multicolumn{4}{c|}{base ($\theta_A = 0$, $\kappa_A = 1$)} & \multicolumn{4}{c}{mandate ($\theta_A = 1$, $\kappa_A = 1$)} \\
\cmidrule(lr){2-5} \cmidrule(lr){6-9}
$\sigma$ & $x_A$ & $x_B$ & $d_A$ & $d_B$ & $x_A$ & $x_B$ & $d_A$ & $d_B$ \\
\midrule
0.0 & 8108 & 203 & 3590 & 995 & 4518 & 203 & 3590 & 995 \\
0.3 & 5615 & 203 & 3590 & 995 & 2025 & 203 & 3590 & 995 \\
0.5 & 3953 & 203 & 3590 & 995 & 363 & 203 & 3590 & 995 \\
0.7 & 2291 & 203 & 3590 & 995 & 0 & 0 & 2493 & 985 \\
0.9 & 628 & 203 & 3590 & 995 & 0 & 0 & 831 & 985 \\
\bottomrule
\end{tabular}
\end{table}

\begin{figure}[htbp]
    \centering
    \includegraphics[width=\textwidth]{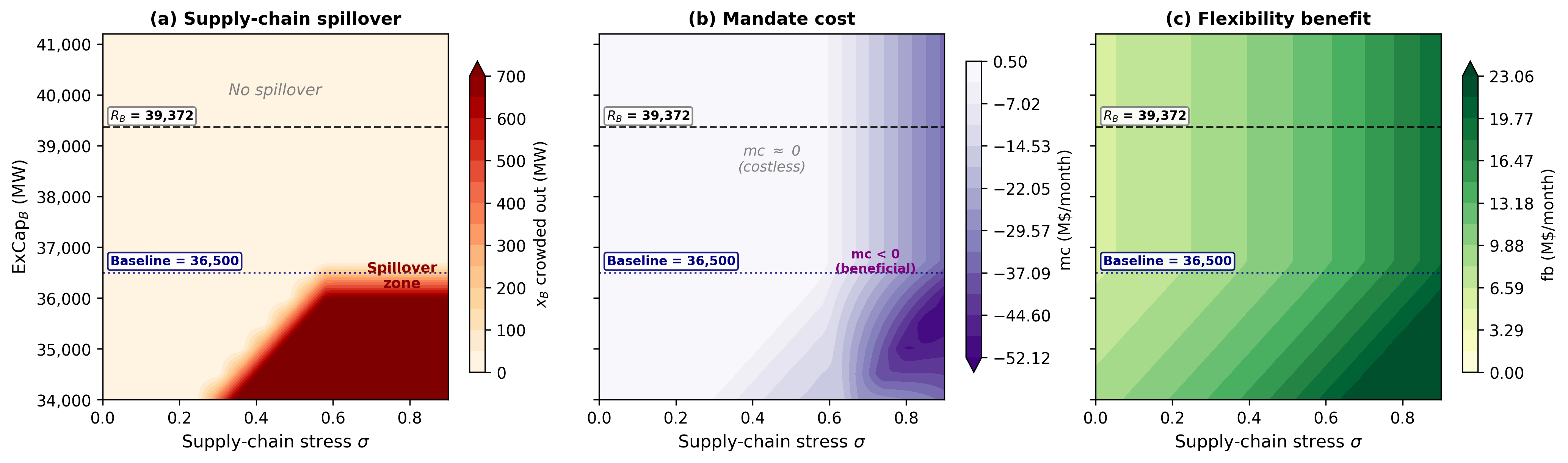}
    \caption{\textbf{Phase diagram of cross-market spillover, mandate cost, and flexibility benefit.} Panels (a) and (b) cover the BYOC mandate (this section); panel (c) previews the flexibility benefit analyzed in Section~\ref{sec:flex}. (a)~Supply-chain crowding out of Market~B entry ($x_B$ lost due to mandate, at $\kappa_A{=}1$). (b)~Mandate cost mc (purple = welfare-improving). (c)~Flexibility benefit fb. Dashed line: $R_B = 39{,}372$~MW. Dotted line: baseline ExCap$_B = 36{,}500$~MW. Spillover occurs at the bottom-right corner of panel~a (high $\sigma$, low ExCap$_B$).}
    \label{fig:phase}
\end{figure}

Figure~\ref{fig:phase} maps the spillover region across the full ($\sigma$, ExCap$_B$) parameter space. Cross-market spillover requires two simultaneous conditions: Market~B must be in deficit (ExCap$_B < R_B$) so the baseline supports new grid entry ($x_B > 0$), and A's mandate must raise $\mu$ enough to crowd out that entry. At lower $\sigma$, the mandate does not raise $\mu$ enough to cross Market~B's entry threshold, so $x_B$ remains unchanged. The spillover region (bottom-right of panel~a) widens as ExCap$_B$ drops or $\sigma$ rises.

The cross-market mechanism is therefore conditional: a unilateral mandate in A can reallocate the fixed equipment budget away from B only when B has active marginal entry. Table~\ref{tab:deficit} shows how mc grows as B's deficit deepens. At the deepest deficit (ExCap$_B = 35{,}000$), the by-market split at $\sigma = 0.7$ shows mc$_A = -59.21$~\$M/mo (gain to A) and mc$_B = +28.54$~\$M/mo (loss to B). A's gain dominates, but B consumers bear significant costs from A's policy.

\begin{table}[H]
\centering\small
\caption{Mandate cost (mc) across the baseline (ExCap$_B = 36{,}500$~MW) and a deeper deficit level (ExCap$_B = 35{,}000$~MW). mc grows in absolute value as B's deficit deepens. The per-market breakdown at the 35,000 level shows Market~B's positive contribution to mc (welfare loss from mandate-induced spillover) rising with deficit depth. All values in \$M/month.}
\label{tab:deficit}
\begin{tabular}{@{}r|rr|>{\columncolor{gray!12}}c >{\columncolor{gray!12}}c@{}}
\toprule
& \multicolumn{2}{c|}{mc (total)} & \multicolumn{2}{c}{mc by mkt at 35,000} \\
\cmidrule(lr){2-3} \cmidrule(lr){4-5}
$\sigma$ & 36,500 (base) & 35,000 (deepest) & A & B \\
\midrule
0.0 & $0$ & $0$ & $0$ & $0$ \\
0.3 & $0$ & $0$ & $0$ & $0$ \\
0.5 & $0$ & $-8.39$ & $\mathbf{-20.92}$ & $\mathbf{+12.53}$ \\
0.7 & $-16.48$ & $\mathbf{-30.67}$ & $\mathbf{-59.21}$ & $\mathbf{+28.54}$ \\
0.9 & $-39.06$ & $\mathbf{-46.27}$ & $\mathbf{-63.58}$ & $\mathbf{+17.30}$ \\
\bottomrule
\end{tabular}
\end{table}

\subsection{Limits of BYOC under non-data-center load growth}
\label{sec:nondc_byoc}

The preceding results assume data centers are the only growing load. In practice, electrification, residential and commercial growth, and other non-DC drivers also increase the reliability requirement; these loads typically cannot participate in BYOC schemes designed for large concentrated loads. To quantify this limitation, we increase $R_A$ by $\Delta R_A \in \{0,\; 2{,}500,\; 5{,}000,\; 10{,}000\}$~MW while holding the DC pipeline ($\bar{d}_A = 5{,}000$~MW) and all other parameters fixed (Table~\ref{tab:nondc}).

\begin{table}[H]
\centering\small
\caption{Mandate-only VRR welfare gap under non-DC load growth. Each cell reports gap $= W^\text{VOLL} - W^\text{VRR}$ at $\theta_A{=}1$, $\kappa_A{=}1$ (mandate without flexibility). $\Delta R_A$ reflects an increase in the reliability requirement, with the DC pipeline held fixed. The $\Delta R_A = 0$ column reproduces the gap column of Table~\ref{tab:mc_vrrvoll}. All values in \$M/month.}
\label{tab:nondc}
\begin{tabular}{@{}rrrrr@{}}
\toprule
& \multicolumn{4}{c}{$\Delta R_A$ (MW)} \\
\cmidrule(lr){2-5}
$\sigma$ & $0$ & $2{,}500$ & $5{,}000$ & $10{,}000$ \\
\midrule
0.0 & $3.06$ & $3.45$ & $12.71$ & \cellcolor{gray!40}$82.47$ \\
0.3 & $3.55$ & $13.25$ & \cellcolor{gray!25}$37.23$ & \cellcolor{gray!55}$170.01$ \\
0.5 & $9.00$ & \cellcolor{gray!25}$27.90$ & \cellcolor{gray!40}$68.24$ & \cellcolor{gray!55}$255.07$ \\
0.7 & $3.91$ & \cellcolor{gray!25}$18.05$ & \cellcolor{gray!40}$53.80$ & \cellcolor{gray!55}$200.26$ \\
0.9 & $1.30$ & $5.54$ & \cellcolor{gray!25}$20.47$ & \cellcolor{gray!40}$76.71$ \\
\bottomrule
\end{tabular}

\vspace{0.5ex}
\parbox{\textwidth}{\footnotesize \textit{Solver note}: in the highest non-DC load cases, Market~A's capacity ratio can enter a low-reserve region ($z_A < 0.925$) where the LP linearization caps VOLL-segment marginal-value coefficients at \$75{,}000/MW-month for numerical conditioning. The affected welfare-gap entries are therefore conservative lower bounds on the true VOLL--VRR welfare difference. The qualitative pattern that the gap increases with non-DC load growth is unaffected.}
\end{table}

Under the mandate alone, the VRR welfare gap grows substantially as non-DC load increases, reaching \$77--\$255~M/mo when $R_A$ grows by 10,000~MW (Table~\ref{tab:nondc}). The mandate is DC-specific: it requires loads that can self-build generation. Flexibility accreditation (Section~\ref{sec:flex}) shares this scope, since it requires loads that can curtail on demand. The table therefore shows that BYOC is a targeted correction for covered large loads, not a general remedy for the reliability and welfare consequences of administrative underpricing as other demand grows.

\section{Flexibility accreditation}
\label{sec:flex}

We now turn to flexibility accreditation ($\kappa_A < 1$), the second policy lever. Unlike the BYOC mandate, which forces self-build to add physical supply, flexibility accreditation reduces the share of data center peak load counted toward the reliability requirement: a data center that contracts to curtail during scarcity is credited as imposing only a fraction $\kappa_A$ of its nominal demand. Section~\ref{sec:welfare_fb} quantifies the aggregate welfare effect and entry response, then Section~\ref{sec:regime} traces a non-monotonic regime transition and its interaction with the BYOC mandate at high stress. Finally, Section~\ref{sec:robustness} isolates the role of supply-chain tightness and benchmarks the gross flexibility gains against illustrative provision costs. Section~\ref{sec:detection} examines whether truthful reporting can be sustained under current penalty rules.

\subsection{Welfare effects and entry response}
\label{sec:welfare_fb}

Table~\ref{tab:part3} first compares the standalone policies, i.e., a full gross-load mandate $(\theta_A,\kappa_A)=(1,1)$ and a 20\% flexibility credit $(0,0.8)$. It then reports the firm-matched package $(0.8,0.8)$ and the conditional comparisons in Eqs.~\eqref{eq:fb_cond}--\eqref{eq:combined}.

\begin{table}[htbp]
\centering\small
\caption{Standalone policy effects and the firm-matched package. mc compares the full gross-load mandate $(1,1)$ with the no-policy case; fb compares $(0,0.8)$ with $(0,1)$. The conditional and combined columns use the firm-matched package $(0.8,0.8)$. All values in \$M/month.}
\label{tab:part3}
\begin{tabular}{@{}rrrrrr@{}}
\toprule
$\sigma$ & mc & fb & fb$|_{\theta_A=0.8}$ & mc$|_{\kappa_A=0.8}$ & combined \\
\midrule
0.0 & $+0.00$ & $+6.38$ & $+6.38$ & $+0.00$ & $+6.38$ \\
0.3 & $+0.00$ & $+9.04$ & $+9.04$ & $+0.00$ & $+9.04$ \\
0.5 & $+0.00$ & $+11.66$ & $+11.66$ & $+0.00$ & $+11.66$ \\
\rowcolor{gray!20} 0.7 & $-16.48$ & $+15.28$ & $+14.48$ & $-9.21$ & $+24.49$ \\
\rowcolor{gray!20} 0.9 & $-39.06$ & $+20.27$ & $+4.63$ & $-23.73$ & $+44.00$ \\
\bottomrule
\end{tabular}
\end{table}

At the $\kappa_A = 0.8$ benchmark, the relative impact on welfare between the two policies is state dependent. For $\sigma \le 0.5$ the mandate is inert (mc $=0$), so flexibility is the only lever. At $\sigma = 0.7$ the two are roughly similar (fb $= 15.28$, $|\text{mc}| = 16.48$); at $\sigma = 0.9$ the mandate is the larger lever (fb $= 20.27$, $|\text{mc}| = 39.06$, shaded rows). Matching the full mandate requires approximately 21\% curtailability ($\kappa_A = 0.785$) at $\sigma = 0.7$ and 37\% curtailability ($\kappa_A = 0.627$) at $\sigma = 0.9$. Appendix~\ref{app:robustness} traces the full $\kappa_A$ dependence.

\begin{figure}[H]
    \centering
    \includegraphics[width=\textwidth]{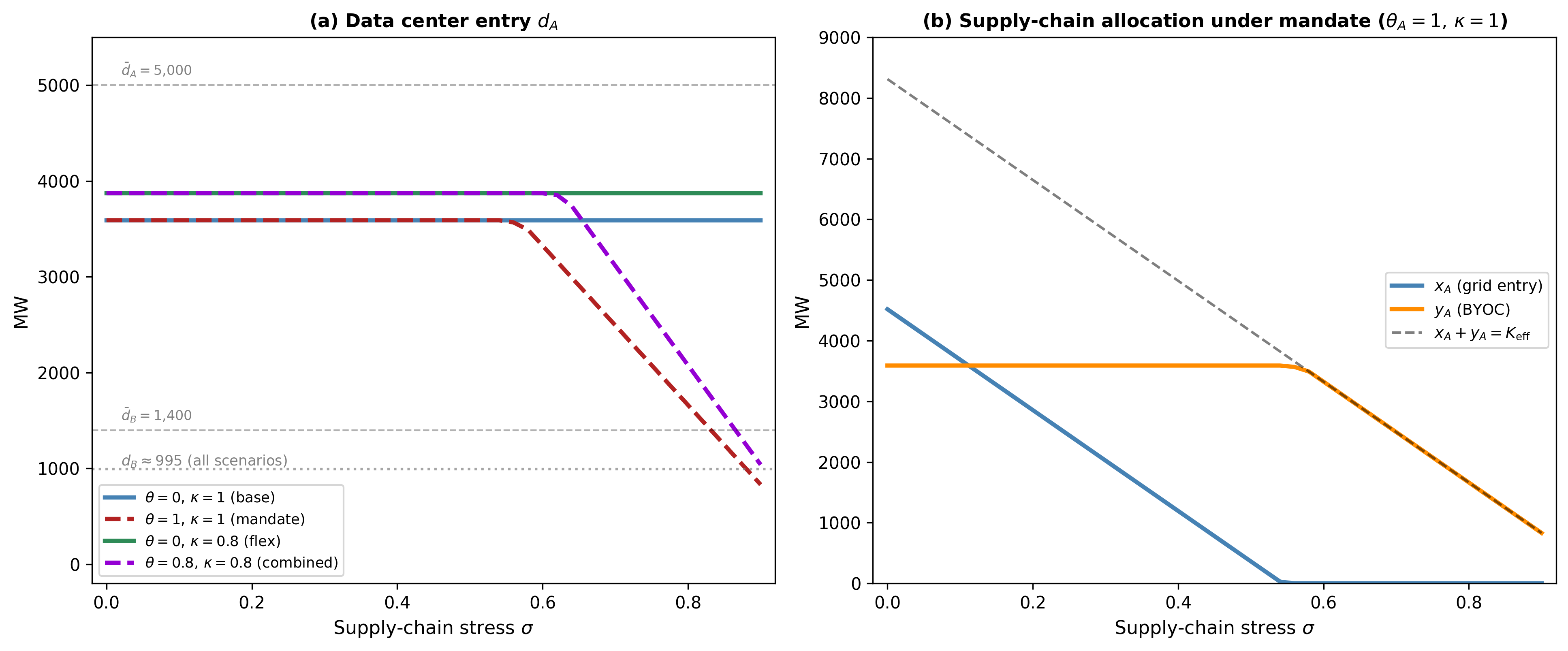}
    \caption{\textbf{Entry and supply-chain allocation across $\sigma$.} (a)~Data center entry $d_A$ under four policy scenarios. At $\kappa_A = 0.8$, entry increases by approximately 282~MW relative to $\kappa_A = 1$ (the flexibility effect). Under the mandate, entry is compressed at high $\sigma$ when the supply chain is exhausted. $d_B \approx 995$~MW across most scenarios, declining slightly to 985~MW under mandate at $\sigma \ge 0.7$. (b)~Supply-chain allocation under mandate ($\theta_A = 1$, $\kappa_A = 1$): grid entry $x_A$ (blue) and BYOC $y_A$ (orange) as $\sigma$ rises. The dashed line shows $K_{\mathrm{eff}} = K_{\mathrm{base}}(1-\sigma)$.}
    \label{fig:entry_supply}
\end{figure}

Combining both instruments yields gross welfare gains of 6 to 44~\$M/month. Figure~\ref{fig:entry_supply} traces how data center entry responds: lowering $\kappa_A$ from 1 to 0.8 raises entry by about 282~MW, while the mandate compresses entry at high $\sigma$ when the supply chain is exhausted.

\subsection{Regime transition and interactions}
\label{sec:regime}

\paragraph{Two-regime mechanism.}\label{sec:two-regime} The conditional benefit fb$|_{\theta_A=0.8}$ in Table~\ref{tab:part3} is non-monotonic: it rises to $+14.48$~\$M ($\sigma=0.7$), then falls to $+4.63$~\$M ($\sigma=0.9$). This non-monotonicity reflects the Regime~1$\to$2 transition introduced in Section~\ref{sec:second_best},  as grid entry is crowded out of the system.

\textbf{Regime 1} ($x_A > 0$, low $\sigma$): Grid entry is positive, so $\gamma = 0$; the DC entry condition~\eqref{eq:dc_foc} then gives $d(\kappa_A) = (v_0 - \kappa_A P) / v_1$. Lowering $\kappa_A$ allows more data center entry ($d$ rises) while simultaneously reducing the net load obligation $\kappa_A d$. Both components improve, yielding $\Delta$Rel $> 0$ and $\Delta$DC $> 0$.

\textbf{Regime 2} ($x_A = 0$, high $\sigma$): Grid entry is fully crowded out. Data center entry is pinned at the remaining supply-chain budget: $\kappa_A$ enters only the capacity position $Q_A$~\eqref{eq:capacity}, not the supply-chain constraint~\eqref{eq:supplychain}, so changing $\kappa_A$ cannot relax the constraint that binds. At the calibrated $K_{\mathrm{base}}$, the binding supply chain also crowds out B's entry ($x_B \to 0$). With $x_A = x_B = 0$, the mandated self-build $y_A = 0.8\,d_A$ absorbs the entire budget, setting $d_A = K_{\mathrm{eff}}/0.8$ with $K_{\mathrm{eff}} = K_{\mathrm{base}}(1-\sigma)$ regardless of $\kappa_A$. Because $d$ cannot respond to changes in $\kappa$, $\Delta$DC $= 0$ and $\Delta$Inv $= 0$. Only $\Delta$Rel remains, driven by the reduced net load of the fixed $K_{\mathrm{eff}}$ capacity. As $\sigma$ increases, $K_{\mathrm{eff}} \to 0$, compressing the remaining consumer reliability gain. Table~\ref{tab:hump_decomp} verifies this decomposition numerically.

\begin{table}[H]
\centering\small
\caption{fb$|_{\theta_A=0.8}$ decomposition for both markets (Eq.~\eqref{eq:fb_decomp}). Both markets shaded by Market~A's regime: lighter for Regime~1 ($x_A > 0$, $\sigma \le 0.5$) and darker for Regime~2 ($x_A = 0$, $\sigma \ge 0.7$). At the $\kappa_A = 0.8$ benchmark, Market~B shows no spillover at any $\sigma$. All values in \$M/month.}
\label{tab:hump_decomp}
\begin{tabular}{@{}crrrr@{}}
\toprule
$\sigma$ & $\Delta$Rel & $\Delta$DC & $\Delta$Inv & fb$|_{\theta_A=0.8}$ \\
\midrule
\multicolumn{5}{@{}l}{\textit{Market A}} \\
\rowcolor{gray!12} 0.0 & $+4.29$ & $+2.10$ & $0$ & $+6.38$ \\
\rowcolor{gray!12} 0.3 & $+6.94$ & $+2.10$ & $0$ & $+9.04$ \\
\rowcolor{gray!12} 0.5 & $+9.56$ & $+2.10$ & $0$ & $+11.66$ \\
\rowcolor{gray!30} 0.7 & $+14.48$ & $0$ & $0$ & $+14.48$ \\
\rowcolor{gray!30} 0.9 & $+4.63$ & $0$ & $0$ & $+4.63$ \\
\addlinespace
\multicolumn{5}{@{}l}{\textit{Market B}} \\
\rowcolor{gray!12} 0.0 & $0$ & $0$ & $0$ & $0$ \\
\rowcolor{gray!12} 0.3 & $0$ & $0$ & $0$ & $0$ \\
\rowcolor{gray!12} 0.5 & $0$ & $0$ & $0$ & $0$ \\
\rowcolor{gray!30} 0.7 & $0$ & $0$ & $0$ & $0$ \\
\rowcolor{gray!30} 0.9 & $0$ & $0$ & $0$ & $0$ \\
\bottomrule
\end{tabular}
\end{table}

At the $\kappa_A = 0.8$ benchmark, Market~B's flexibility spillover is zero at every $\sigma$ (Table~\ref{tab:hump_decomp}): the shallow flexibility credit does not shift B's clearing price or grid entry.

\paragraph{Interactions with BYOC.}\label{sec:interactions_fb}
Table~\ref{tab:part3} shows that the two policy instruments operate independently at low stress but interact at high stress. At $\sigma = 0.3$, the combined welfare effect ($+9.04$~\$M) exactly equals fb$|_{\theta_A=0.8}$ ($+9.04$) plus mc ($0$), yielding zero interaction. At $\sigma = 0.9$, a negative interaction emerges: the unconditional fb reaches $+20.27$~\$M, but conditioning on the mandate (fb$|_{\theta_A=0.8}$) attenuates this to $+4.63$~\$M, an interaction of $-15.63$~\$M (defined as $\text{fb}|_{\theta_A=0.8} - \text{fb}$). This attenuation occurs because the mandate limits the entry channel that flexibility operates on. Without a mandate, lower $\kappa_A$ allows more data center entry ($d_A$ rises from 3,590 to 3,872~MW) while reducing net load ($\kappa_A d_A$ falls from 3,590 to 3,098~MW), generating both private value and reliability gains. Under the mandate, this entry channel remains open in Regime~1 where $x_A > 0$ and flexibility still unlocks new $d_A$, but closes in Regime~2 where entry is pinned by the supply chain and the private value gain vanishes (Figure~\ref{fig:entry_supply}).

\subsection{Drivers and caveats}
\label{sec:robustness}

\paragraph{Supply-chain tightness.}\label{sec:slack}
Both the mandate's second-best value and the growth of fb with $\sigma$ come from the binding supply chain. To show this, we raise $K_{\mathrm{base}}$ until the constraint no longer binds ($\mu = 0$ at every $\sigma$). In this abundant-equipment counterfactual, the supply chain constraint is slack, mc $= 0$ at every $\sigma$, and fb is constant at \$4.6M/month. Thus, the second-best value of BYOC and the stress-dependence of flexibility are conditional on physical equipment scarcity. The hard-cap specification also implies that, once the constraint binds, administrative pricing primarily changes load entry and the allocation of a fixed quantity of equipment rather than aggregate equipment deployment.

\paragraph{Illustrative flexibility provision cost.}\label{sec:provision}
The reported fb is a gross benefit. We do not include the private cost of shifting compute, operating backup resources, testing, telemetry, or compliance in the equilibrium. As an illustrative opportunity cost range, we multiply flexible capacity $(1-\kappa_A)\,d_A$ by an annual curtailment duration of 7--35 hours~\citep{Camus_FlexibleDataCenters_2025} and a compute value of \$2{,}100--6{,}000/MWh (Mainstream Modern AI, interruptible service class)~\citep{Billimoria_Byers_VoCL_2026}. Compute value is a market price rather than a social surplus estimate, so this calculation should be interpreted as an illustrative private opportunity cost rather than a welfare estimate. The resulting range runs from \$1.0M/month at the low anchor to \$13.5M/month at the high anchor. Subtracting the proxy against fb, the difference stays positive at every $\sigma$ under the low anchor; under the high anchor it turns negative for $\sigma \le 0.5$ but stays positive at high stress, where fb is largest. As such, the low anchor leaves the state-dependent ranking intact, while the high anchor makes fb minus the proxy negative at low stress.

\subsection{Incentive compatibility of flexibility reporting}
\label{sec:detection}

At the 20\% flexibility-credit benchmark, the welfare gain from flexibility may not be visible in the clearing price. When the VRR cap binds in Market~A ($\sigma \ge 0.3$ in our calibration), $P_A$ stays at its administrative limit regardless of $\kappa_A$.

This lack of visibility raises the question of whether data centers will report flexibility truthfully when prices alone do not penalize misreporting.

We focus on a firm-matched design in which the BYOC coverage share adjusts with the reported firm share. Assume that a data center knows its true firm-share requirement $\kappa_\text{true}$ but reports $\hat{\kappa} \in [0,\kappa_\text{true}]$ to the system operator. A lower report represents a claim of greater
flexibility and reduces both the facility's ordinary capacity obligation
and its BYOC coverage requirement. Further, reporting $\hat{\kappa} < \kappa_\text{true}$ exposes the data center to a penalty $f$ per MW of shortfall if a scarcity event occurs in a given month (with probability $p$).

For the following partial-equilibrium calculation we assume an individual data center takes $P$, $\gamma$, and the penalty rule as given when choosing its reported flexibility. The analysis applies to any market $m$ with a capacity price $P$ and penalty rate $f$; we drop the market subscript for clarity.

Conditional on its reported $\hat{\kappa}$, the data center
chooses entry to solve
\begin{equation}
    \Pi(\hat{\kappa})
    =
    \max_d
    \left\{
        V(d)
        -
        \hat{\kappa}(P+\gamma)d
        -
        p f
        \left(
            \kappa_{\mathrm{true}}-\hat{\kappa}
        \right)d
    \right\}.
    \label{eq:profit}
\end{equation}
The associated entry condition is
\begin{equation}
    V'(d^*)
    =
    \hat{\kappa}(P+\gamma)
    +
    p f
    \left(
        \kappa_{\mathrm{true}}-\hat{\kappa}
    \right).
    \label{eq:entry_reporting}
\end{equation}
Applying the envelope theorem gives
\begin{equation}
    \frac{d\Pi}{d\hat{\kappa}}
    =
    d^*
    \left[
        p f-(P+\gamma)
    \right].
    \label{eq:envelope}
\end{equation}
Truthful reporting is locally incentive-compatible against downward
deviations when
\begin{equation}
    P+\gamma \leq p f.
    \label{eq:ic}
\end{equation}
Using the cost-conservation identity,
\[
    P+\gamma=C+\mu,
\]
the condition can equivalently be written as
\begin{equation}
    C+\mu \leq p f.
    \label{eq:ic_cost}
\end{equation}
Thus, the relevant private benefit from overstating flexibility is the
constrained marginal cost of capacity, not only the auction price
visible to regulators and market monitors.

Equation~\eqref{eq:ic_cost} can fail because either the expected frequency of enforcement or the penalty rate is too low.
In PJM, non-performance is charged at $C_\text{yr}/30$ per hour of a Performance Assessment Interval (PAI), where $C_\text{yr} = 12C_A$ is the annual Net CONE~\citep{PJM_Manual18_VRR}. Meanwhile, PJM’s 2026/27 resource-adequacy analysis estimates 0.54 expected PAI hours per delivery year~\citep{PJM_PAI}. With $C_A = \$5,500/\text{MW-mo}$ and using the PJM forecast as an illustrative comparison, this formula leads to an expected penalty of $0.54 \cdot 12\cdot C_A/30 = \$1{,}188/\text{MW-yr}$. At the modeled cap, the visible capacity cost alone is $12\cdot P_A = \$99,000/\text{MW-yr}$, making the expected charge about 1.2\% of the visible price saving. When $\gamma>0$, the mismatch grows further because the data center avoids $P+\gamma=C+\mu$, while the penalty does not increase with the hidden mandate premium.
This conclusion follows from treating data centers as strategic agents responding to financial penalties alone. In practice, data centers have additional incentives to report flexibility truthfully: reduced grid reliability degrades their own operations, and reputation effects, compliance norms, and verification mechanisms support alignment through channels outside our model. The broader policy challenge is therefore not simply escalating penalties but combining aligned price signals with these complementary channels.

\section{Discussion}
\label{sec:discussion}

This paper compares BYOC mandates and flexibility accreditation in capacity markets that share a constrained equipment supply chain. Under efficient pricing and the model's assumption of perfect substitution, requiring data centers to procure capacity mainly reallocates between grid-built and BYOC capacity and has little welfare effect. When an administrative demand curve caps the visible capacity price below the constrained marginal cost \(C+\mu\), however, BYOC can become a second-best correction in the short run. The mandate makes covered loads internalize the equipment scarcity that is missing from the auction price and thereby disciplines data center entry. The correction is targeted, applying only to loads that can procure or self-build accredited capacity, so it does not eliminate the welfare gap associated with growth by residential, commercial, and other loads outside the mandate. A unilateral mandate can also generate cross-market consequences, depending on the degree of equipment scarcity and the neighbor's deficit depth.

At the same time, the static welfare gain from BYOC does not make it a general solution to problems in the capacity market. Higher capacity prices increase payments to the full incumbent capacity fleet. BYOC preserves much of the reduction in load payments created by an administrative cap while imposing the missing marginal cost on selected new entrants through a separate obligation. The combination of price caps and BYOC could further weaken the market in the long run if it reinforces the expectation that future demand growth and scarcity rents will be carved out through bilateral or administrative procurement. Given the consumer protection motivation for these interventions, a more durable approach to reform could be to introduce stronger ex ante hedging mechanisms that would protect customers from price spikes while preserving the underlying price signal~\cite{Shu2023}.

Flexibility accreditation acts on a different margin by reducing the capacity obligation itself. At low and moderate supply-chain stress, only flexibility is a welfare-improving instrument; the effect of BYOC becomes comparable or larger than flexibility only under severe stress. Larger flexibility credits shift the ranking toward flexibility, but its net value also depends on the cost of provision.
Furthermore, translating these benefits into practice requires credible enforcement. If PJM's current non-performance penalty design were adapted to a firm-matched flexible-load product, it would not align incentives for truthful reporting, suggesting that additional verification mechanisms beyond financial penalties are needed to align reporting incentives.
For data centers, offering verifiable flexibility in exchange for faster grid access can be mutually beneficial under correctly aligned incentives, since reduced grid reliability also degrades data center operations.

Several limitations qualify these results. The framework is static, two-market, and single-technology. It omits  retirement dynamics and bilateral capacity obligations across interconnected regions, as well as intrazonal details such as downstate New York bottlenecks~\citep{NYISO_RNA_2024}. Supply-chain constraints enter through a single aggregate parameter~\citep{Gorman_Joule_2024, Yao_SupplyChain_2025}. Because equipment availability is represented by a hard upper bound, once that bound binds the policies primarily alter load entry and the allocation of a fixed quantity of equipment rather than aggregate equipment deployment. Our calibration also treats NYISO as a policy-neutral baseline, while FERC's June 2026 show-cause order~\citep{FERC_LargeLoadShowCause_2026} makes BYOC-adjacent provisions a live possibility not only in Market~B but across all six U.S. RTOs/ISOs. We also take the flexibility parameters as exogenous and measure the gross benefits provided by flexibility rather than calculating the cost to provide flexibility.  Extending the model to  heterogeneous technologies, upward-sloping equipment supply, dynamic investment and retirement, and explicit flexibility contracts would allow these assumptions to be relaxed.

The central distinction is nevertheless robust: BYOC changes who internalizes the cost of capacity, whereas flexibility changes the capacity obligation itself. Administrative price caps determine when BYOC becomes a useful second-best correction, while physical equipment scarcity and the cost of credible curtailment determine the value of flexibility. As these interventions move from proposals to tariff filings, the trade-offs quantified here
are the terms on which the new rules will be judged.

\bibliographystyle{unsrtnat}
\bibliography{references}

\appendix

\section{VOLL demand curve derivation}
\label{app:voll_derivation}

This appendix presents the full derivation underlying the calibration
of $\mathrm{VOLL}_m$ and $\beta_m$ in Section~\ref{sec:calibration}.

\subsection{VOLL level from Net CONE}
\label{app:voll_level}

We follow the implied-VOLL framework of~\citet{Zhao_Zheng_Litvinov_2018}.
The annualized demand curve at the system level equals VOLL times the
marginal reduction in expected unserved energy (EUE) caused by adding
capacity:
\begin{equation}
D_{\mathrm{annual},m}(z) = \mathrm{VOLL}_m \cdot \left| \frac{\partial \mathrm{EUE}_m}{\partial Q_m} \right|.
\end{equation}
For installed capacity (ICAP), \citet[eq.~36]{Zhao_Zheng_Litvinov_2018}
establishes the reliability identity
\begin{equation}
-\frac{\partial \mathrm{EUE}_m}{\partial Q_m} = \mathrm{LOLH}_m(z) \cdot (1 - \mathrm{EFORd}_m),
\end{equation}
where $\mathrm{EFORd}_m$ is the demand-weighted equivalent forced outage
rate.
Anchoring the curve at the reliability requirement ($z=1$) by
forcing the annualized marginal value to equal the annual Net CONE,
\begin{equation}
D_{\mathrm{annual},m}(1) = 12\, C_m,
\end{equation}
gives the closed-form expression
\begin{equation}
\mathrm{VOLL}_m = \frac{12\, C_m}{\mathrm{LOLH}_{0,m} \cdot (1 - \mathrm{EFORd}_m)}.
\label{eq:voll_appendix}
\end{equation}

We set $\mathrm{EFORd}_m = 4.1\%$ for both markets, matching the derating
factor for new fossil peaking plants from NERC GADS data on units no
more than ten years old~\citep[p.~57]{NYISO_DCR_2024, NERC_GADS}. This
value reflects new-build reliability rather than the fleet average
(e.g., PJM's UCAP/ICAP ratio of $0.7969$), which includes aging and
intermittent resources. With $\mathrm{LOLH}_{0,A} = 0.323$~h/yr from
PJM's 2025/26 BRA~\citep{PJM_PC_3IA_2025} and
$\mathrm{LOLH}_{0,B} = 0.374$~h/yr from NYISO's installed capacity
requirement~\citep{NYSRC_Peak_IRM}, Eq.~\eqref{eq:voll_appendix} yields
$\mathrm{VOLL}_A = \$213{,}070$/MWh and $\mathrm{VOLL}_B = \$191{,}375$/MWh.
For external validation, \citet[eq.~37]{Zhao_Zheng_Litvinov_2018} applies
the same formula to ISO-NE and obtains $\$216{,}048$/MWh, within 1.4\%
of our PJM estimate.

\subsection{Slope $\beta$ calibration}
\label{app:beta}

The slope $\beta_m$ is the exponential decay rate of LOLE with respect
to the normalized reserve margin $z = Q/R$:
\begin{equation}
\mathrm{LOLE}_m(z) = \mathrm{LOLE}_{0,m} \exp(-\beta_m (z-1)),
\qquad \mathrm{LOLE}_{0,m} = 0.1 \text{ events/yr.}
\label{eq:lole_exp}
\end{equation}
The reference value $\mathrm{LOLE}_{0,m} = 0.1$ corresponds to the
1-in-10 reliability standard adopted by both PJM and NYISO. We calibrate
$\beta_m$ from different presentations of the same underlying
LOLE--reserve-margin relationship in the two markets.

\paragraph{NYISO: forward-MRI method.}
Differentiating Eq.~\eqref{eq:lole_exp} and evaluating at $z=1$ gives\\
$|d\mathrm{LOLE}/dQ|_{z=1} = \beta_m \mathrm{LOLE}_0 / R_m$. Identifying
this with the marginal reliability impact (MRI) reported in
\citet[Fig.~29]{Potomac_SOM_2024}, we solve for $\beta_B$:
\begin{equation}
\beta_B = \frac{\mathrm{MRI} \cdot R_B}{\mathrm{LOLE}_0}
= \frac{(0.0052/100) \cdot 34{,}059}{0.1} = 17.7.
\end{equation}
Here the MRI is converted from days per 100~MW to days per MW,
$R_B = 34{,}059$~MW is the NYISO UCAP-basis installed capacity
requirement, and $\mathrm{LOLE}_0 = 0.1$ days/yr is the standard target.
The choice of UCAP versus ICAP basis cancels in the ratio
$\mathrm{MRI} \cdot R_B$, so $\beta_B$ is dimensionless and basis-invariant.

\paragraph{PJM: curve-fit method.}
\citet{Brattle_VRR_2025} provides the LOLE--reserve-margin
curve from PJM's PRISM reliability model across
$z \in [0.96, 1.06]$. Taking logarithms of Eq.~\eqref{eq:lole_exp}
linearizes the exponential form, $\ln \mathrm{LOLE} = \ln \mathrm{LOLE}_0 - \beta_m (z-1)$,
and fitting this relation to the published curve yields
$\beta_A \approx 35$. The fit is approximately exponential, with
mild asymmetry: the curve is slightly steeper below the reliability
requirement than above~\citep{Brattle_VRR_2025}.

\subsection{LOLE--LOLH approximation and sensitivity}
\label{app:lole_lolh}

The demand-curve formula uses LOLH, while our $\beta$ estimates come
from LOLE data. The two indices differ by an event-duration factor:
$\mathrm{LOLH} = \mathrm{LOLE} \cdot D$, where $D(z)$ is the average
loss-of-load event duration in hours. Because outage events tend to
grow longer as reserve margins tighten, $D(z)$ also decays with $z$,
so $\beta_{\mathrm{LOLH}} > \beta_{\mathrm{LOLE}}$.

\section{Sensitivity to market calibration}
\label{app:shapley}

We use a Shapley decomposition of $4! = 24$ permutations to quantify the sensitivity of the flexibility benefit (fb, Equation~\ref{eq:fb}) and mandate cost (mc, Equation~\ref{eq:mc}) to Market~B's calibration (Table~\ref{tab:shapley}). From a symmetric baseline, we switch four parameters to their Market~B calibration values: $C_B$ ($5{,}500 \to 5{,}720$), $\beta_B$ ($35.0 \to 17.7$), $\bar{d}_B$ ($5{,}000 \to 1{,}400$), and ExCap$_B$ ($0.933 R_B \to 36{,}500$), holding structural rules ($R_B = 39{,}372$, ICAP curve) fixed. These Shapley values isolate shifts in policy metrics rather than absolute welfare. For example, fixing $\bar{d}_B = 1{,}400$~MW, we recompute fb (varying $\kappa_A$ from 1.0 to 0.8) and mc (varying $\theta_A$ from 0 to 1), attributing the resulting change directly to that parameter switch.

\begin{table}[H]
\centering\small
\caption{Shapley decomposition of B-market calibration effects on fb and mc by $\sigma$. Shaded cells: individual $|\text{contribution}| \ge 4$~\$M. All values in \$M/month.}
\label{tab:shapley}
\begin{tabular}{@{}lrrrrr@{}}
\toprule
& $\sigma{=}0.0$ & $\sigma{=}0.3$ & $\sigma{=}0.5$ & $\sigma{=}0.7$ & $\sigma{=}0.9$ \\
\midrule
fb$|_{\text{sym}}$ & $+6.22$ & $+8.77$ & $+11.29$ & $+14.77$ & $+19.57$ \\
$+\;C_B$ & $-0.01$ & $-0.13$ & $-0.18$ & $-0.25$ & $-0.35$ \\
$+\;\beta_B$ & $-0.04$ & $+0.00$ & $+0.00$ & $+0.00$ & $-0.00$ \\
$+\;\bar{d}_B$ & $+0.23$ & $+0.26$ & $+0.35$ & $+0.49$ & $+0.67$ \\
$+\;\text{ExCap}_B$ & $-0.00$ & $+0.14$ & $+0.20$ & $+0.27$ & $+0.37$ \\
$=$ fb$|_{\text{cal}}$ & $+6.38$ & $+9.04$ & $+11.66$ & $+15.28$ & $+20.27$ \\
$\Delta$fb & $+0.16$ & $+0.27$ & $+0.37$ & $+0.51$ & $+0.70$ \\
\addlinespace
mc$|_{\text{sym}}$ & $+0.00$ & $+0.00$ & $+0.00$ & $-14.12$ & $-34.59$ \\
$+\;C_B$ & $+0.00$ & $+0.00$ & $-0.00$ & $+0.73$ & $+1.79$ \\
$+\;\beta_B$ & $+0.00$ & $+0.00$ & $+0.00$ & $-2.62$ & $-2.62$ \\
$+\;\bar{d}_B$ & $+0.00$ & $+0.00$ & $+0.00$ & $-0.18$ & $-2.22$ \\
$+\;\text{ExCap}_B$ & $+0.00$ & $+0.00$ & $-0.00$ & $-0.28$ & $-1.41$ \\
$=$ mc$|_{\text{cal}}$ & $+0.00$ & $+0.00$ & $+0.00$ & $-16.48$ & $-39.06$ \\
$\Delta$mc & $0$ & $0$ & $0$ & $-2.36$ & $-4.47$ \\
\bottomrule
\end{tabular}
\end{table}

For fb, the contributions are small at every $\sigma$: $\bar{d}_B$ is the largest single term ($+0.23$ to $+0.67$), and the total shift $\Delta$fb runs from $+0.16$ to $+0.70$~\$M. The flexibility benefit is robust to the Market~B calibration. For mc, all contributions are zero at $\sigma \le 0.5$; at $\sigma \ge 0.7$, the dominant effect is $\beta_B$ ($-2.62$), giving $\Delta$mc of $-2.36$ to $-4.47$~\$M---i.e., the NYISO-scale calibration deepens the mandate's welfare improvement at high stress.

\section{Sensitivity to data center value and flexibility}
\label{app:robustness}

This appendix tests the sensitivity of data center entry and the flexibility benefit to the data center marginal entry value $v_0$, which the main text holds fixed at the calibrated value.

Table~\ref{tab:v0_sweep} reports the policy metrics across $v_0 \in \{20{,}000, 25{,}000, 29{,}260, 35{,}000, 40{,}000\}$~\$/MW-month at the baseline supply-chain stress $\sigma = 0.3$. The calibrated value $v_0 = 29{,}260$ sits in the middle of the range.

\begin{table}[H]
\centering\small
\caption{Policy metrics across $v_0$ at $\sigma = 0.3$ (unilateral, $\theta_B = 0$, $\kappa_B = 1$). All welfare values in \$M/month.}
\label{tab:v0_sweep}
\begin{tabular}{@{}rrrrrrr@{}}
\toprule
$v_0$ & $d_A$ (MW) & mc & fb & fb$|_{\theta_A=0.8}$ & combined & interact \\
\midrule
20{,}000 & 2{,}938 & $+0.00$ & $+6.22$ & $+6.22$ & $+6.22$ & $+0.00$ \\
25{,}000 & 3{,}350 & $+0.00$ & $+7.89$ & $+7.89$ & $+7.89$ & $+0.00$ \\
29{,}260 & 3{,}590 & $+0.00$ & $+9.04$ & $+9.04$ & $+9.04$ & $+0.00$ \\
35{,}000 & 3{,}822 & $+0.00$ & $+10.29$ & $+10.29$ & $+10.29$ & $+0.00$ \\
40{,}000 & 3{,}970 & $+0.00$ & $+11.16$ & $+11.16$ & $+11.16$ & $+0.00$ \\
\bottomrule
\end{tabular}
\end{table}

Data center entry $d_A$ grows monotonically with $v_0$. The mandate cost mc remains zero across the entire range because $\sigma = 0.3$ keeps the system in Regime~1 ($x_A > 0$, $\gamma_A = 0$), where the mandate is a pure relabeling regardless of $v_0$. The flexibility benefit fb scales roughly linearly with $v_0$ (from $+6.22$ at $v_0 = 20{,}000$ to $+11.16$~\$M/mo at $v_0 = 40{,}000$; the qualitative pattern, mc $= 0$ and fb $> 0$ at $\kappa_A = 0.8$, holds at every $v_0$ tested). We focus on $\sigma = 0.3$ as the calibration anchor (where mc $= 0$ is structural and $v_0$ only shifts $d_A$ and fb in magnitude); at higher $\sigma$, the second-best correction mechanism in Section~\ref{sec:second_best} is driven by the widening gap between the capped price and the efficient price $C_A + \mu$ rather than by $v_0$, and the regime transition analysis in Section~\ref{sec:regime} characterizes the resulting non-monotonicity directly.

\subsection*{Flexibility credit}

Table~\ref{tab:kappa_sweep} varies $\kappa_A$ across the full stress range.

\begin{table}[H]
\centering\small
\caption{Flexibility benefit fb and ratio fb$/|\text{mc}|$ across $\kappa_A$ and $\sigma$ (unilateral). mc $= 0$ at $\sigma \le 0.5$, $-16.48$ at $\sigma = 0.7$, $-39.06$ at $\sigma = 0.9$. Shaded: flexibility does not dominate. fb in \$M/month.}
\label{tab:kappa_sweep}
\begin{tabular}{@{}r rrrr | rr@{}}
\toprule
& \multicolumn{4}{c|}{fb (\$M/month)} & \multicolumn{2}{c}{fb$/|\text{mc}|$} \\
$\kappa_A$ & $\sigma{=}0.3$ & $\sigma{=}0.5$ & $\sigma{=}0.7$ & $\sigma{=}0.9$ & $\sigma{=}0.7$ & $\sigma{=}0.9$ \\
\midrule
0.2 & $+36.21$ & $+47.80$ & $+63.78$ & $+85.81$ & $3.87$ & $2.20$ \\
0.4 & $+27.53$ & $+36.09$ & $+47.90$ & $+64.18$ & $2.91$ & $1.64$ \\
0.6 & $+18.34$ & $+23.87$ & $+31.49$ & $+42.00$ & $1.91$ & $1.08$ \\
0.8 & $+9.04$  & $+11.66$ & $+15.28$ & $+20.27$ & \cellcolor{gray!25}$0.93$ & \cellcolor{gray!25}$0.52$ \\
\bottomrule
\end{tabular}
\end{table}

At the main-text $\kappa_A = 0.8$ benchmark, flexibility exceeds the full mandate for $\sigma \le 0.5$ but not at $\sigma \ge 0.7$ (fb$/|\text{mc}| = 0.93$ and $0.52$, shaded). Larger flexibility credits raise the ratio monotonically, with flexibility exceeding the mandate at every $\sigma$ once $\kappa_A \le 0.6$.

\end{document}